\documentclass[a4paper,fleqn,usenatbib]{mnras}

\usepackage{newtxtext,newtxmath}

\usepackage[T1]{fontenc}
\usepackage{ae,aecompl}

\usepackage{graphicx}	
\usepackage{amsmath}	
\usepackage{amssymb}	
\usepackage{bm}
\usepackage{comment}

\newcommand{\dd}{\,\mathrm{d}}
\newcommand{\au}{\,\mathrm{au}}
\newcommand{\BV}{Br\"unt-V\"ais\"al\"a }
\newcommand{\lO}[0]{_{\mathrm{O}}}
\newcommand{\lH}[0]{_{\mathrm{H}}}
\newcommand{\lA}[0]{_{\mathrm{A}}}
\newcommand{\lR}[0]{_{\mathrm{R}}}
\newcommand{\Tbb}[0]{T_{\mathrm{bb}}}
\newcommand{\Tg}[0]{T_{\mathrm{g}}}
\newcommand{\Ttau}[0]{T_{\tau}}
\newcommand{\Byt}[0]{B_y^{\rm top}}
\newcommand{\Bxt}[0]{B_x^{\rm top}}

\graphicspath{ {Pics/} }

\title[Electric heating in laminar protoplanetary disks]{Electric heating and angular momentum transport in laminar models of protoplanetary disks}

\author[W. B\'ethune \& H. Latter]{
William B\'ethune$^{1}$\thanks{E-mail: william.bethune@uni-tuebingen.de} and Henrik Latter$^{2}$
\\
$^{1}$ Institut f\"ur Astronomie und Astrophysik,
Universit\"at T\"ubingen, Auf der Morgenstelle 10, 72076 T\"ubingen, Germany\\
$^{2}$ DAMTP, University of Cambridge, CMS, Wilberforce Road, Cambridge, CB3 0WA, UK\\
}

\date{Accepted XXX. Received YYY; in original form ZZZ}

\pubyear{2020}

\begin{document}
\label{firstpage}
\pagerange{\pageref{firstpage}--\pageref{lastpage}}
\maketitle

\begin{abstract}
  The vertical temperature structure of a protoplanetary disk bears on several processes relevant to planet formation, such as gas and dust grain chemistry, ice lines and convection. The temperature profile is controlled by irradiation from the central star and by any internal source of heat such as might arise from gas accretion. We investigate the heat and angular momentum transport generated by the resistive dissipation of magnetic fields in laminar disks. We use local one-dimensional simulations to obtain vertical temperature profiles for typical conditions in the inner disk (0.5 to 4 au). Using simple assumptions for the gas ionization and opacity, the heating and cooling rates are computed self-consistently in the framework of radiative non-ideal magnetohydrodynamics. We characterize steady solutions that are symmetric about the midplane and which may be associated with saturated Hall-shear unstable modes. We also examine the dissipation of electric currents driven by global accretion-ejection structures. In both cases we obtain significant heating for a sufficiently high opacity. Strong magnetic fields can induce an order-unity temperature increase in the disk midplane, a convectively unstable entropy profile, and a surface emissivity equivalent to a viscous heating of $\alpha \sim 10^{-2}$. These results show how magnetic fields may drive efficient accretion and heating in weakly ionized disks where turbulence might be inefficient, at least for a range of radii and ages of the disk. 
\end{abstract}



\section{Introduction}

The imaging of dust substructures \citep{alma15,andrews18} and measurements of gas kinematics \citep{flaherty15,flaherty17,louvet18} have provided valuable clues to the long-standing issues of gas accretion and planet formation in circumstellar disks. Of particular relevance for planet formation is the thermal structure of the disk, which is tightly connected to the accretion process. 

The temperature distribution determines the radial location of ice lines, with consequences on grain growth \citep{lorek18}, the preferential sites of planet formation \citep{ida2008,cridland17,hyodo19} and their composition \citep{dodson09,cridland16,bitsch19}. The temperature of the disk controls its chemistry \citep{walsh15,kamp2017}, which goes into our interpretation of molecular emission lines \citep[e.g.,][]{flaherty15,oya2019}. Finally, by affecting the ionization state of the gas \citep{weingartner01,ilgner12}, the temperature may also influence how magnetic fields impact on the disk dynamics.

The temperature of the disk results from a balance between stellar irradiation, internal heating mechanisms and radiative cooling. In the inner regions of protoplanetary disks, heating is dominated by the liberation of gravitational energy through gas accretion \citep{alessio98}. Magnetic fields are believed to play a major role in the transport of angular momentum responsible for accretion, whether they power magnetized outflows \citep{pudritznorman83,pelletierpudritz92} or seed turbulence via the magnetorotational instability \citep[MRI,][]{balbushawley91,hawleybalbus92}. However, the details of angular momentum transport and energy dissipation are often concealed in an effective viscosity parameter $\alpha$ \citep{shakusunyaev73}. The bulk accretion rates are consistent with $\alpha \sim 10^{-4} - 10^{-2}$ in young stellar objects \citep{belllin94}, but the accretion process might be vertically inhomogeneous \citep{gammie96}, and linking $\alpha$ to a vertical temperature profile pre-supposes that the accretion energy is dissipated locally \citep{balbuspap99}. 

When accounting for the low ionization fraction, magnetohydrodynamic (MHD) simulations support a picture of laminar accretion in the inner regions of protoplanetary disks. Angular momentum is transported by a non-turbulent magnetic stress through the disk \citep{LKF14,bai2015} and by magneto-thermal winds out of the disk \citep{bethune17,bai2017}. Given the complex nature of global MHD simulations, only a few studies have incorporated radiative effects \citep{wang19,rodenkirch20} with a focus on the disk-wind interaction. Electric dissipation is predicted to efficiently heat the wind \citep{safier93a}, but its contribution inside the disk was only recently examined by \citet{mori+19} who concluded that it should be negligible when compared to stellar irradiation. However, the isothermal model of \citet{mori+19} did not self-consistently solve for the thermal structure of the disk.

We compute the one-dimensional (1D) vertical structure of the inner $0.5-4\au$ in radiative MHD models of protoplanetary disks. Our model improves on previous studies by self-consistently describing the energy exchanges due to Ohmic resistivity, the Hall effect, ambipolar diffusion and radiative transport for plausible ionization fractions and opacities. A high-order spatial method helps us resolve the sharp current sheets developing in ambipolar MHD \citep{brandenburg94} and our method is free from the constraints of explicit time integration schemes, allowing us to probe resistivity regimes inaccessible to common simulation codes. 

We link the resistive electric heating to the accretion power and characterize them in terms of a gas temperature and an effective `viscosity' coefficient $\alpha$. We consider two different drivers of mass accretion / electric currents inside the disk. In the first case, electric currents are amplified by the Hall-shear instability \citep[HSI,][]{kunz08} and we examine the 1D saturated phase of this instability. Because the instability extracts orbital energy from inside the disk, we designate these solutions as \emph{internally-driven}. In the second case, we impose the total electric current passing through the disk and let the current density find the path of least resistance. This situation mimics magnetized accretion-ejection for which a proper energy budget would involve the entire disk-wind system. Since we focus on the internal structure of the disk only, the accretion power appears to be \emph{externally-driven}.

In both cases, the resistive dissipation of electric currents can generate as much heat as viscous disks models with $\alpha \sim 10^{-4} - 10^{-2}$, and correspondingly large mass accretion rates. If the disk is opaque at thermal wavelengths, this heat can build up in the midplane and dominate over stellar irradiation in the thermal balance of the disk.

We present our disk model and the system of radiative MHD equations describing it in Sect. \ref{sec:method}. We also detail the theoretical framework in which our results can be interpreted. The results are split into two sections depending on the origin of the accretion power: we present internally-driven solutions in Sect. \ref{sec:mri} and externally-driven solutions in Sect. \ref{sec:ext}. We interpret these results and discuss the limitations of our model in Sect. \ref{sec:discussion} before concluding. 

\section{Method} \label{sec:method}

The thermal structure of a passive disk is solely governed by the stellar irradiation, its reprocessing and re-emission at thermal wavelengths. The dissipation of electric currents introduces an additional source of heat in magnetized disks. We aim to compute the vertical structure of axisymmetric protoplanetary disks threaded by a net poloidal magnetic field, irradiated by the central star, and subject to such non-ideal MHD processes.

We consider quasi-equilibrium structures whose relaxation times are short compared to the long-time and large-scale evolution of the disk via mass losses, magnetic flux transport and the disk's changing radiative environment. We also assume that at any given disk radius, the main physics determining these states is independent of neighboring radial annuli. Given these restrictions, the most natural framework is the stratified shearing box whose input parameters (irradiation flux, surface density, etc.) depend on radius according to a pre-defined global disk model. 

\subsection{Model and governing equations}

\subsubsection{Global disk model} \label{sec:diskmodel}

At a distance $r$ from a Sun-like star of radius $R_{\odot}=6.957\times 10^{10}\,\mathrm{cm}$ and effective temperature $T_{\odot}=5777\,\mathrm{K}$, the equilibrium black-body temperature of a disk is:
\begin{equation} \label{eqn:deftbb}
  \Tbb = T_{\odot} \left(\frac{R_{\odot}}{r}\right)^{1/2} \left(\frac{h}{r}\right)^{1/4},
\end{equation}
depending on the passive opening angle of the disk
\begin{equation}
  \frac{h}{r} = \left( \frac{k_{\mathrm{B}} T_{\odot} \sqrt{R_{\odot} r}}{\mu m\lH G M_{\odot}} \right)^{4/7},
\end{equation}
see for example \cite{dullemond2000}. In these equations, $k_{\mathrm{B}}$ is Boltzmann's constant, $\mu=2.353$ is the mean molecular weight of the gas for a prescribed solar composition, $m_{\mathrm{H}}$ the hydrogen mass and $M_{\odot}=1.989\times 10^{33} \,\mathrm{g}$ the mass of the Sun. We take solar parameters for simplicity, noting that representative protostars might have larger radii and lower effective temperatures. 

We prescribe the mass surface density of the disk $\Sigma = 1.7 \times 10^3 \left(r/1\mathrm{au}\right)^{-3/2} \mathrm{g}\,\mathrm{cm}^{2}$ consistently with the Minimum Mass Solar Nebula \citep{hayashi81} to facilitate comparisons with previous works. Although recent surveys point toward shallower density profiles $\Sigma \sim r^{-1/2}$ \citep{andrews07a,tazzari17}, using a steeper density profile allows us to sample a broader range of disk conditions and thus cover the uncertainties in disk masses and ages \citep{berginwilliams17}. 

\subsubsection{Stratified shearing sheet}

At a chosen radius $r$ around the star, we move into a reference frame orbiting at the local orbital frequency $\Omega$ and expand the potential of the star to second order following the `shearing sheet' approximation \citep{goldreich65,hill78,latterpap17}. We obtain the vertical structure of the disk at this radius by integrating the equations of radiative MHD in the shearing sheet.

We describe the gas as a mixture of neutrals, ions, and electrons, with the neutrals dominating the gas mass and the free electrons being the mobile charge carriers. Denoting by $\rho$ the neutral gas density, $\bm{u}$ its velocity, and $\bm{B} = B \bm{e_B}$ the magnetic (induction) field, their evolution in time is governed by:
\begin{align}
D_t \rho &= -\rho\nabla\cdot\bm{u}, \label{eqn:DTrho}\\
D_t \bm{u} &= -\frac{1}{\rho}\nabla P -2\Omega \bm{e}_z\times\bm{u}+ 2q\Omega^2 x\bm{e}_x -\Omega_z^2 z \bm{e}_z -\frac{\bm{J}\times\bm{B}}{\rho}, \label{eqn:DTu}\\
D_t \bm{B} &= \bm{B}\cdot\nabla \bm{u}+\nabla\times \bm{\mathcal{E}}, \label{eqn:DTB}
\end{align}
where we introduced the operator $D_t=\partial_t+\bm{v}\cdot\nabla$, the shear rate $q=-d\log\Omega/d\log r$ and the vertical epicyclic frequency $\Omega_z$. In \eqref{eqn:DTB}, $\bm{\mathcal{E}}$ is the electric field in a frame comoving with the neutral fluid (detailed in Sect. \ref{sec:conductivity}) and $\bm{J}=\nabla\times\bm{B}$ is the electric current neglecting relativistic effects. 

For the gas pressure $P$ we consider an ideal diatomic gas with a single adiabatic index $\gamma=7/5$. Denoting $E\lR$ the frequency-integrated radiation energy density, the gas pressure and radiation energy are coupled via
\begin{align}
D_t P &= -\gamma P \nabla\cdot \bm{u}
         -(\gamma-1)\frac{c}{\lambda}(a \Tg^4-E\lR)
         +(\gamma-1)Q, \label{eqn:DTP}\\
\partial_t E\lR &= -\nabla\cdot \bm{F}\lR + \frac{c}{\lambda}(a \Tg^4-E\lR), \label{eqn:DTer}
\end{align}
where $c$ is the speed of light, $\lambda$ is the photon mean free path, $a\equiv 4\sigma_{\mathrm{SB}}/c$ with $\sigma_{\mathrm{SB}}$ the Stefan-Boldzmann constant, $\Tg=P/\rho$ is the gas temperature, $Q$ is the heating power density of dissipative effects and $\bm{F}\lR$ is the flux of radiative energy (detailed in Sect. \ref{sec:radiation}). 

We consider axisymmetric equilibria that vary on global scales radially. In the shearing sheet these radial variations are neglected and thus our local solutions depend only on the vertical coordinate --- pointing along the axis of rotation and denoted by $z$. Although the radial coordinate (usually denoted by $x$) disappears in the final form of the equations, the radial shear of the flow still affects the momentum and magnetic induction equations.

Furthermore, we neglect the radial pressure gradient in the disk by considering that it orbits the star at the Keplerian frequency\footnote{The deviations from Keplerian velocity scale as $h/r$ relative to the sound speed, i.e., at most $3.7\times 10^{-2}$ in the model considered here, see Table \ref{tab:disk}. We also neglect the vertical shear $\partial_z u_y$ induced by the radial temperature profile \eqref{eqn:deftbb} of the passive disk \citep[e.g.,][]{urpin84}.} for every $z$. In this case the shear rate $q=3/2$ and $\Omega_z=\Omega$. The shearing-sheet equations \eqref{eqn:DTrho}-\eqref{eqn:DTu} then support the steady isothermal solution of a Keplerian shear flow $\partial_x u_y = -\left(3/2\right) \Omega$. 

Despite these simplifications, the problem retains complexity in the details of the electromotive field $\mathcal{E}$ and radiative energy flux $F\lR$. We describe these terms in the following paragraphs and give the final form of the radiative MHD equations in Sect. \ref{sec:equations}. 

\subsubsection{Ionization fraction} \label{sec:ionfraction}

Solving for the detailed chemical composition of the gas in a dusty environment is expensive in computational time and subject to strong assumptions. We use the same simplifications as \cite{LKF14} to evolve the ionization fraction $x_e$ in time as a balance between the local ionization versus recombination rates. In particular, we consider a metal and dust-free environment of primordial chemical composition ($75\%$ hydrogen).

The ionization rate includes contributions from stellar X-rays \citep{igeaglass99} using the fit of \citet[][first term of equation 4]{gressel15}, cosmic rays with a penetration depth of $96\,\mathrm{g}\,\mathrm{cm}^{-2}$ \citep{umebayanakano81}, and radioactive decay at a constant ionization rate of $10^{-19}\,\mathrm{s}^{-1}$ \citep{umebayanakano09}. The X and cosmic ray fluxes are assumed to penetrate the disk vertically from both sides. We neglect collisional ionization at the temperatures $<10^3\,\mathrm{K}$ considered.

Including dust grains and/or metals can alter the ionization fraction by several orders of magnitude depending on their abundance and distribution in the disk \citep[e.g.,][]{sano2000,fromang2002,wardle07}. A simple way to account for the dust-enhanced recombination rate is to artificially reduce the ionization fraction. We therefore include models with an ionization fraction $x_e$ reduced by a factor $10^{-2}$ in Sect. \ref{sec:lowion} and $10^{-3}$ in Sect. \ref{sec:ext}. 

\begin{figure}
\begin{center}
\includegraphics[width=\columnwidth]{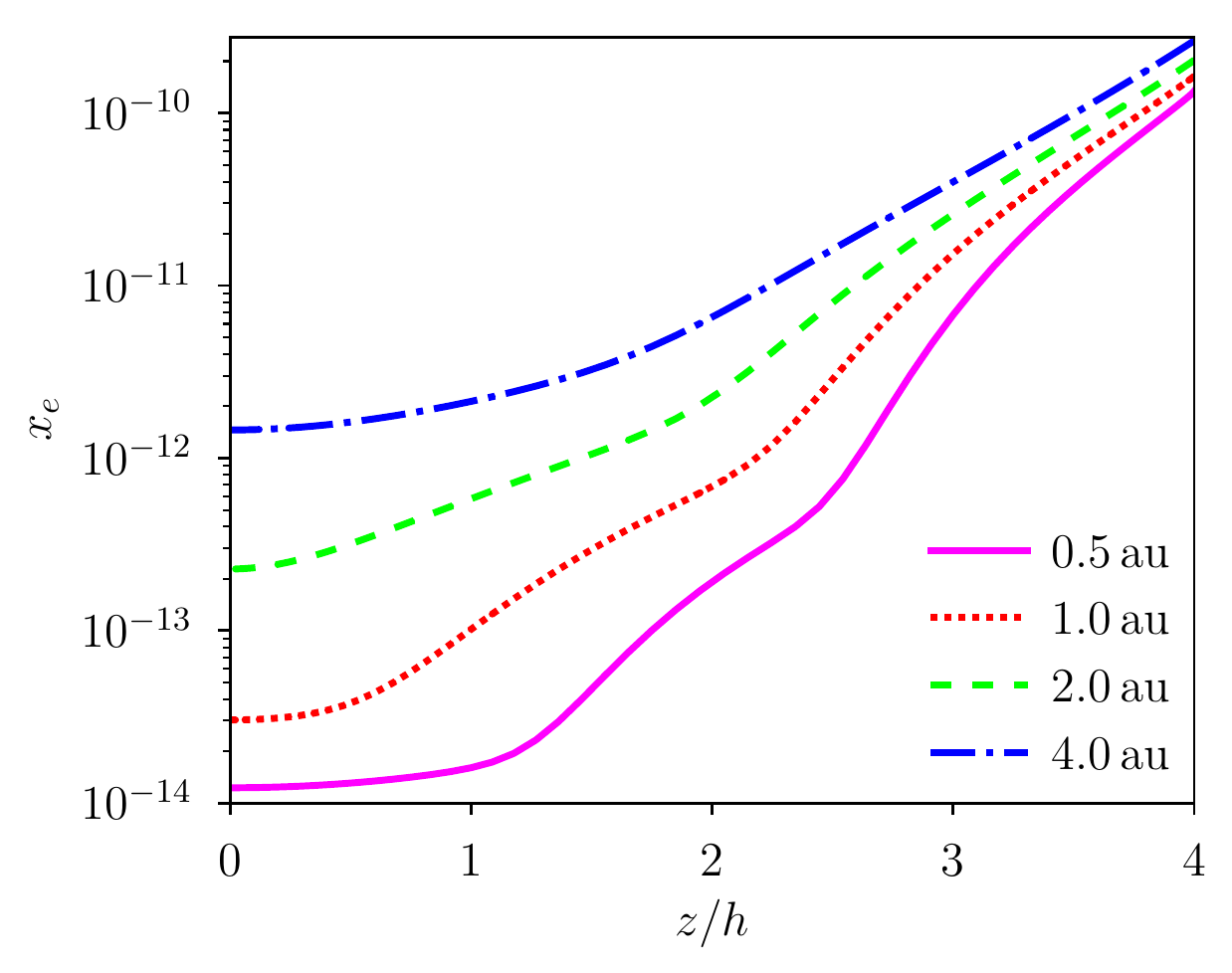}
\caption{Vertical profiles of ionization fraction $x_e$ for an isothermal disk of temperature $\Tbb$ given by \eqref{eqn:deftbb} at different disk radii (see legend). \label{fig:ionfraction}}
\end{center}
\end{figure}

Far ultra-violet (FUV) stellar radiations can ionize carbon and sulfur in the uppermost layers of the disk, providing a floor value $x_e \geq 10^{-5}$ down to column densities $\lesssim 10^{-2} \,\mathrm{g}\,\mathrm{cm}^{-2}$ \citep[$z/h \approx 4$,][]{perezchiang11}. We do not include this source of ionization because it favors variability in the uppermost layers of the disk and hinders convergence to steady states \citep{riols16}. The ionization profiles corresponding to the reference model described above are drawn on Fig. \ref{fig:ionfraction}.

\subsubsection{Electric conductivity} \label{sec:conductivity}

At such low ionization fractions $x_e \lesssim 10^{-10}$, the plasma imperfectly conducts electric currents. In the frame co-moving with the neutral gas, the electric field can be related to the electric current by a generalized Ohm's law:
\begin{equation}
  \bm{\mathcal{E}} = - \underbrace{\bm{u} \times \bm{B}}_{\rm ideal} + \underbrace{\eta\lO \bm{J}}_{\rm Ohm} + \underbrace{\eta\lH \bm{J}\times\bm{e_B}}_{\rm Hall} - \underbrace{\eta\lA \left(\bm{J}\times\bm{e_B}\right)\times\bm{e_B}}_{\rm ambipolar},
\end{equation}
The `non-ideal' diffusivities $\eta_{\rm O,H,A}$ are evolved as in \citet{LKF14} assuming that the plasma is composed of neutrals, electrons and ions; we do not include dust grains as charge carriers. 

\begin{figure}
\begin{center}
\includegraphics[width=\columnwidth]{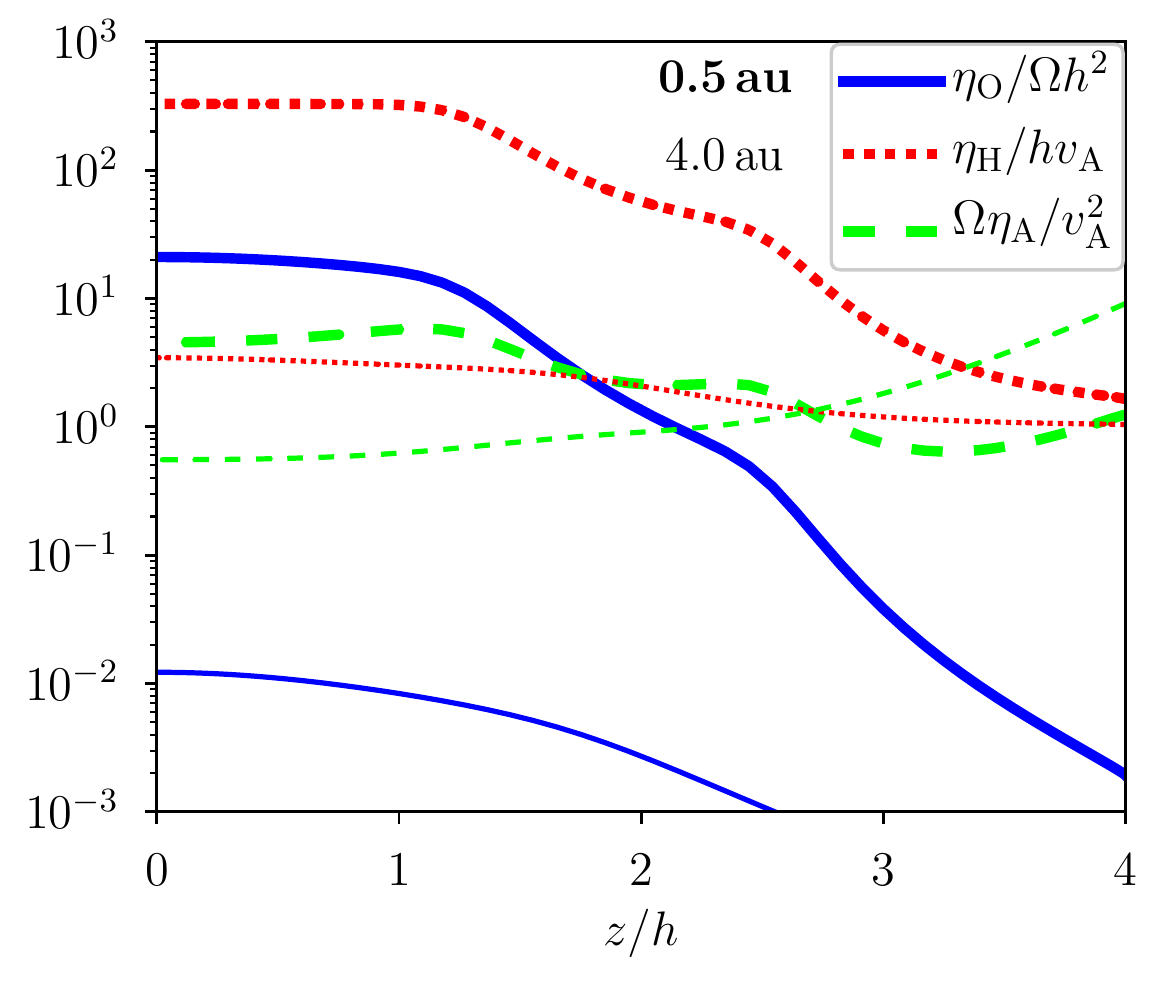}
\caption{Vertical profiles of dimensionless numbers characterizing the strength of Ohmic (solid blue), Hall (dotted red) and ambipolar (dashed green) diffusivities for a vertically isothermal disk at $0.5\au$ (thick lines) and $4.0\au$ (thin lines) with the ionization fractions shown on Fig. \ref{fig:ionfraction}. \label{fig:nonideal}}
\end{center}
\end{figure}

The relative importance of the three non-ideal MHD effects can be characterized by appropriately normalizing the diffusivities $\eta_{\rm O,H,A}$. Let $v\lA \equiv B / \sqrt{\rho}$ denote the Alfv\'en velocity. The vertical profiles of $\eta\lO/\Omega h^2$, $\eta\lH/h v\lA$ and $\Omega \eta\lA/v\lA^2$ are drawn on Fig. \ref{fig:nonideal} at a distance of $0.5\au$ and $4\au$ from the star in the passive and vertically isothermal disk described in Sect. \ref{sec:diskmodel}. These dimensionless numbers are independent of the strength of the magnetic field and can be used to determine the linear stability of the disk to the MRI (see Sect. \ref{sec:linstab} and references therein).

\subsubsection{Radiative energy flux} \label{sec:radiation}

We solve for the frequency-integrated radiation energy density $E\lR$ in the flux-limited-diffusion (FLD) approximation. We compute the frequency-integrated opacity $\kappa$ as a function of the local gas temperature $\Tg$ and neutral gas density $\rho$ using the tables of \citet{belllin94}. The photon mean free path is $\lambda \equiv 1/\kappa \rho$ and the optical depth $\tau\left(z\right) \equiv -\int_{\infty}^z \kappa \rho \dd z'$ is integrated toward the disk midplane.

In the FLD approximation, the radiative energy flux is
\begin{equation}
  F\lR = - f_{\mathrm{M}} c \lambda \frac{\partial E\lR}{\partial z}. \label{eqn:radflux}
\end{equation}
where the flux limiter $f_{\mathrm{M}}$ of \citet{minerbo78} allows a smooth transition from the optically thick regime $F\lR \simeq - \left(\lambda c / 3\right) \partial_z E\lR$ to the optically thin regime $F\lR \simeq - c E\lR \left( \partial_z E\lR \,/ \left\vert\partial_z E\lR\right\vert\right)$. The radiation pressure $E\lR/3$ is typically $10^{-6}$ times smaller than the gas pressure in the regime considered, so we neglect momentum exchanges between the gas and radiation field. 

The stellar irradiation is incorporated in the problem by imposing the radiation temperature $T\lR \equiv \sqrt[4]{E\lR / a}$ at low optical depth above the disk. This approach provides the correct black-body temperature in the optically thick parts of a passive disk while ignoring the details of heat deposition in its surface layers. 

\subsubsection{Governing equations} \label{sec:equations}

Solving for the deviations from the background Keplerian shear $\bm{v} = \bm{u} - \left(3/2\right)\Omega x \bm{e}_y$, we look for steady states of the following system of equations:
\begin{align}
  \frac{\partial \rho}{\partial t} &= -\frac{\partial}{\partial z}\left( \rho v_z\right),\label{eqn:dtrho}\\
  \frac{\partial v_x}{\partial t} &= 2\Omega v_y - v_z \frac{\partial v_x}{\partial z} + \frac{J_y B_z}{\rho}, \label{eqn:dtvx}\\
  \frac{\partial v_y}{\partial t} &= -\frac{1}{2}\Omega v_x - v_z\frac{\partial v_y}{\partial z} - \frac{J_x B_z}{\rho}, \label{eqn:dtvy}\\
  \frac{\partial v_z}{\partial t} &= -\frac{\partial}{\partial z} \left(\frac{1}{2}v_z^2+\frac{1}{2}\Omega^2 z^2\right) - \frac{1}{\rho} \frac{\partial}{\partial z} \left[ P + \frac{1}{2} \left(B_x^2 + B_y^2\right) \right] \nonumber\\
  &+ \Omega h^2 \partial_z^2 v_z, \label{eqn:dtvz}\\
  \frac{\partial B_x}{\partial t} &= \frac{\partial \mathcal{E}_y}{\partial z}, \label{eqn:dtbx}\\
  \frac{\partial B_y}{\partial t} &= -\frac{\partial \mathcal{E}_x}{\partial z} - \frac{3}{2}\Omega B_x, \label{eqn:dtby}\\
  \frac{\partial P}{\partial t} &= -v_z\frac{\partial P}{\partial z} - \gamma P \frac{\partial v_z}{\partial z} - \left(\gamma-1\right) \frac{c}{\lambda} \left(a \Tg^4 - E\lR\right) \nonumber\\
  &+\left(\gamma-1\right) \left[\eta_O \left(J_x^2+J_y^2\right) + \eta_A \left(J_{\perp x}^2+J_{\perp y}^2+J_{\perp z}^2\right)\right], \label{eqn:dtp}\\
  \frac{\partial E\lR}{\partial t} &= -\frac{\partial F\lR}{\partial z} + \frac{c}{\lambda} \left(a \Tg^4 - E\lR\right). \label{eqn:dter}
\end{align}
The last term in \eqref{eqn:dtvz} acts as a viscosity to damp vertical motions, allowing the disk to relax to an equilibrium. In \eqref{eqn:dtp}, ambipolar heating appears as a function of the electric current projected perpendicularly to the local magnetic field: $\bm{J_{\perp}} = -\left(\bm{J}\times\bm{e}_B\right)\times\bm{e}_B = \bm{J} - \left(\bm{J}\cdot\bm{e_B}\right) \bm{e_B}$.

\subsubsection{Units and conventions}

We can identify a set of natural scales in this problem. The orbital frequency $\Omega$ is taken as inverse time unit. The passive scale height $h$ of the disk --- without internal heating --- is taken as distance unit. The velocity unit is therefore the passive sound speed $c_s=\Omega h$. Note that the actual density stratification scale varies and becomes larger than $h$ when the temperature increases inside the disk. The gas temperature $\Tg\equiv P/\rho$ is normalized by the surface black-body value $\Tbb$. Finally, the gas surface density $\Sigma$ defines a mass unit. 

The degree of magnetization of the disk is measured by the dimensionless parameter
\begin{equation} \label{eqn:defbeta}
  \beta \equiv \frac{\Sigma \Omega^2 h}{B_z^2}.
\end{equation}
For an isothermal hydrostatic equilibrium, $\beta$ is approximately $20$ per cent larger than the midplane ratio of thermal versus magnetic pressures $\beta_0 \equiv 2 \rho c_s^2 / B_z^2$. 

\autoref{tab:disk} gathers typical values of our disk model at different radii. The choice of disk radius strongly affects the surface density of the gas and only weakly its black-body temperature. In turn, the surface density controls the MHD diffusivities and the gas opacity. 

\begin{table}
\begin{center}
\caption{Characteristics of the chosen disk model at different radii: gas surface density $\Sigma$, passive opening angle $h/r$, black-body disk temperature $\Tbb$, altitude $z_{\tau}$ of the $\tau=1$ surface, vertically averaged opacity $\overline{\kappa}$. \label{tab:disk}}
\begin{tabular}{cccccc}
$r/\au$ & $\Sigma / \mathrm{g}\,\mathrm{cm}^{-2} $ & $h/r$ & $\Tbb / \mathrm{K}$ & $z_{\mathrm{\tau}}/h$ & $\overline{\kappa} / \mathrm{cm}^2\,\mathrm{g}^{-1}$\\
  \hline
  $0.5$ & $4.81 \times 10^3$ & $2.04\times 10^{-2}$ & $211$ & $3.58$ & $1.47$\\
  $1.0$ & $1.70 \times 10^3$ & $2.49\times 10^{-2}$ & $156$ & $3.58$ & $4.27$\\
  $2.0$ & $6.01 \times 10^2$ & $3.03\times 10^{-2}$ & $116$ & $3.22$ & $2.70$\\
  $4.0$ & $2.13 \times 10^2$ & $3.70\times 10^{-2}$ & $86$ & $2.73$ & $1.49$\\
\end{tabular}
\end{center}
\end{table}

\subsection{Numerics}

We obtain the steady-state vertical structure of the disk by an initial value approach. We integrate the equations \eqref{eqn:dtrho}-\eqref{eqn:dter} in time until a steady-state criterion is satisfied. This method does not require a good guess of the solution to start with, and it guarantees that the steady-state solutions are stable to disturbances of the flow variables that only depend on the vertical direction. 

\subsubsection{Computational domain}

We only solve the equations on the upper half of the disk ($z \ge 0$), and assume that our solutions exhibit an equatorial symmetry about the midplane. This choice helps reduce computational costs while allowing control of the midplane conditions to machine accuracy.

The vertical domain is fixed to $z/h\in \left[0,4\right]$ throughout this paper. As long as the effective scale height of the disk is approximately $h$, most of the gas mass and electric current are located inside the computational domain. This domain also encloses the $\tau=1$ altitude, so it captures the transition from the optically thin upper regions ($\lambda/h\gtrsim 1$) to the optically thick midplane ($\lambda/h\ll1$). Convergence with domain size is discussed in Appendix \ref{app:convtest}. 

The domain is meshed with $64$ Gauss-Lobatto points to perform Chebyshev differentiation and integration by simple matrix-vector products. This spectral decomposition provides the maximal accuracy for smooth solutions at the cost of numerical resilience when the flow variables exhibit sharp gradients. 

\subsubsection{Integration scheme}

The scale separation between the radiative time, MHD diffusive time and sound-crossing time makes the problem computationally inaccessible to purely explicit integration schemes. We therefore integrate the equations via a fully implicit scheme for all the flow variables. Because our interest is in steady states we can adopt first order integration without concerns over time-accuracy. We demonstrate in Appendix \ref{app:integrator} that this numerical scheme does capture the growth of MRI modes both in space and time. 

We rescale the flow variables to have similar amplitudes and cast the system of equations \eqref{eqn:dtrho}-\eqref{eqn:dter} in the form $\partial_t f - S(f) = 0$. We stop the integration when a solution satisfies the simple steady-state criterion $\| \left( \partial_t f \right) / \Omega f \|_{\infty} < 10^{-4}$. The residual error generally keeps converging to zero when this criterion is satisfied. 

\subsubsection{Initial and boundary conditions} \label{sec:initbound}

We start the time integration close to the `current-free' equilibrium
\begin{equation} \label{eqn:current-free}
  \begin{pmatrix}
    \rho\\
    v_x, v_y, v_z\\
    B_x, B_y, B_z\\
    P\\
    E\lR
  \end{pmatrix}
  \left(z,t=0\right)
  =
  \begin{pmatrix}
    \rho_0 \exp\left(-z^2 / 2h^2\right)\\
    0,\, 0,\, 0\\
    0,\, 0,\, \beta^{-1/2}\\
    \rho_0 c_s^2 \exp\left(-z^2 / 2h^2\right)\\
    a \Tbb^4
  \end{pmatrix},
\end{equation}
consisting of the isothermal Keplerian flow with a vertical magnetic field $B_z\neq 0$ in it. The density is normalized so that $\int_0^{4h} \rho \dd z = \Sigma / 2$.

On top of this equilibrium, the horizontal velocity and magnetic field components are initialized with random noise $\left(\tilde{v}_x,\tilde{v}_y,\tilde{B}_x,\tilde{B}_y\right)$ of amplitude $10^{-4}$. To speed-up parametric explorations, a previously computed solution is re-used as the initial condition if only one control parameter has changed since. 

We enforce an equatorial symmetry at the midplane via
\begin{equation} \label{eqn:lowbound}
  \left(
  \partial_z \rho,\,
  \partial_z v_x,\,
  \partial_z v_y,\,
  v_z,\,
  B_x,\,
  B_y,\,
  \partial_z P,\,
  \partial_z E\lR
  \right)\left(z=0,t\right) = 0.
\end{equation}
These conditions will pick out equilibria that exhibit `hourglass' magnetic configurations through the midplane. We consider two possible sets of boundary conditions at the top of the domain. The first set is
\begin{equation} \label{eqn:topbound}
  \begin{pmatrix}
    v_z,\,
    \partial_z B_x,\,
    \partial_z B_y,\,\\
    E\lR
  \end{pmatrix}
  \left(z=4h,t\right)
  =
  \begin{pmatrix}
    0,\,
    0,\,
    0\,\\
    a \Tbb^4
  \end{pmatrix}
\end{equation}
while we let the other flow variables relax to stationary values. Note that \eqref{eqn:topbound} does not impose the orientation nor strength of the magnetic field, which the system select itself. This set of conditions will be associated with `internally-driven' equilibria in Sect. \ref{sec:mri}. The second set of boundary conditions differs from \eqref{eqn:topbound} only by imposing the value $\Byt \equiv B_y(4h) \neq 0$, and will be associated with `externally-driven' states in Sect. \ref{sec:ext}.

\subsection{Diagnostics} \label{sec:diagnostics}

We measure the gas temperatures $T_0$ at the midplane and $\Ttau$ at the $\tau=1$ altitude. We define the temperature contrast as $\left(T_0-\Ttau\right)/\Ttau$. We define the specific entropy $s \equiv P / \rho^{\gamma}$ and deduce the squared \BV frequency:
\begin{equation} \label{eqn:bvn2}
  \mathcal{N}^2 \equiv \frac{\Omega^2 z}{\gamma s} \frac{\partial s}{\partial z}.
\end{equation}
Negative values of $\mathcal{N}^2<0$ imply convective instability in a purely hydrodynamic disk with no viscosity or thermal diffusion \citep{ruden88,held18}, and assuming that non-ideal MHD effects erase any stabilisation from magnetic tension. We define the \BV growth rate $\omega \equiv \sqrt{-\mathcal{N}^2}$ when $\mathcal{N}^2<0$ and zero otherwise. 

The surface value of the azimuthal magnetic field $\Byt$ can be related to the net electric current passing through the disk via
\begin{equation} \label{eqn:byjx}
  B_y^{\mathrm{top}} = \int_{0}^{4h} \!\!\!\!\!\!-J_x \dd z,
\end{equation}
and to the mass accretion rate $\dot{M} = 2\int_0^{4h} \rho v_x \dd z $ after multiplying \eqref{eqn:dtvy} by $\rho$ and integrating:
\begin{equation} \label{eqn:stressacc}
  B_y^{\mathrm{top}} = \frac{\Omega \dot{M}}{4 B_z}.
\end{equation}
This last equation connects the net mass accretion rate to the angular momentum extracted vertically by the magnetic stress $-B_y^{\mathrm{top}} B_z$. 

In steady state, the condition $v_z=0$ at the boundaries enforce $v_z=0$ everywhere, i.e., no advective flux of kinetic or thermal energy through the domain. The pressure equation \eqref{eqn:dtp} then becomes a competition between Ohmic and ambipolar heating versus radiative cooling. To measure their influence on the gas temperature, we introduce the respective heating/cooling rates per unit mass:
\begin{equation} \label{eqn:qheat}
\partial_t P = \rho \left(q\lO + q\lA + q\lR\right) = 0,
\end{equation}
where $q\lO$ denotes Ohmic heating, $q\lA$ ambipolar heating and $q\lR$ radiative cooling. We also define the total energy density
\begin{equation} \label{eqn:defe}
E \equiv \frac{1}{2}\rho \bm{v}\cdot\bm{v} + \frac{1}{2} \bm{B}\cdot\bm{B} + \frac{1}{\gamma-1} P + E\lR. 
\end{equation}
If $\bm{L}\equiv \bm{J}\times\bm{B}$ is the Lorentz force and $v_z=0$ because of boundary conditions, then the evolution of the total energy density \eqref{eqn:defe} obeys
\begin{alignat}{2} \label{eqn:nrj}
  \partial_t E = &+\frac{3}{2}\Omega \left( \rho v_x v_y - B_x B_y \right) &\mathrm{sources} \nonumber\\
  &- \partial_z \left[\left(v_x B_x + v_y B_y\right)B_z \right] &\mathrm{ideal} \nonumber\\
  &-\partial_z \left[\left(J_x B_y - J_y B_x\right) \eta\lO\right] &\mathrm{Ohmic}\\
  &-\partial_z \left[\left(L_x B_y - L_y B_x\right) \eta\lH \right] &\mathrm{Hall} \nonumber\\
  &-\partial_z \left[\left(J_{\perp x} B_y - J_{\perp y} B_x\right) \eta\lA \right] \qquad &\mathrm{ambipolar} \nonumber\\
  &-\partial_z F\lR &\mathrm{radiation.} \nonumber
\end{alignat}
It comprises a single source term, arising from the extraction of orbital energy by the combined action of Reynolds and Maxwell stresses. The subsequent terms represent the ideal MHD Poynting flux, three energy fluxes due to non-ideal MHD effects and the radiative energy flux. As emphasized throughout this paper, the four fluxes of magnetic energy are thermodynamically crucial because they can redistribute orbital energy away from the height at which it was originally extracted, and before this energy is be thermalized.

The heat generated by electron-neutral (Ohmic) and ion-neutral (ambipolar) collisions is integrated vertically to define the electric heat fluxes per unit surface of the disk:
\begin{align} 
  Q\lO &\equiv \int_{0}^{4h} \eta\lO \left(J_x^2+J_y^2\right) \dd z = \int_{0}^{4h} \frac{\rho q\lO}{\gamma-1} \dd z, \label{eqn:heat_ohm}\\
  Q\lA &\equiv \int_{0}^{4h} \eta\lA \left(J_{\perp x}^2+J_{\perp y}^2+J_{\perp z}^2\right) \dd z = \int_{0}^{4h} \frac{\rho q\lA}{\gamma-1} \dd z. \label{eqn:heat_amb}
\end{align}

Integrating \eqref{eqn:nrj} vertically and substituting \eqref{eqn:dtp} \& \eqref{eqn:dter} in steady state, we can connect the heat fluxes with the radial flux of angular momentum through the disk:
\begin{equation} \label{eqn:momheat}
\frac{3}{2}\Omega \int_0^{4h} \left( \rho v_x v_y - B_x B_y \right) \dd z = Q\lO+Q\lA + F_z^{\mathrm{top}},
\end{equation}
where $F_z^{\mathrm{top}}$ is the magnetic energy flux (ideal, Ohmic, Hall and ambipolar) through the upper boundary $z=4h$.

Normalizing the heat fluxes into a dissipation coefficient
\begin{equation}\label{eqn:defalpha}
  \alpha \equiv \frac{8}{9} \frac{Q\lO + Q\lA}{\Sigma h^2 \Omega^3},
\end{equation}
we can relate stress and dissipation in the standard framework of $\alpha$ disks \citep{balbuspap99}. When no magnetic energy flows through the upper boundary, $F_z^{\mathrm{top}} = 0$ and we have
\begin{equation} \label{eqn:alphadisk}
  \int_0^{4h} \left( \rho v_x v_y - B_x B_y \right) \dd z = \frac{3}{4} \alpha \Sigma h^2 \Omega^2.
\end{equation}
On the other hand, it the energy extracted by the internal $xy$ stress is negligible compared to the energy flux $F_z^{\mathrm{top}}$ at the surface of the disk, we obtain the balance:
\begin{equation} \label{eqn:alphatorque}
  F_z^{\mathrm{top}} = - \frac{9}{8} \alpha \Sigma h^2 \Omega^3.
\end{equation}

\section{Internally-driven states} \label{sec:mri}

\subsection{Instability of the current-free equilibrium} \label{sec:linstab}

Integrating the system \eqref{eqn:dtrho}-\eqref{eqn:dter} in time subject to the boundary conditions \eqref{eqn:topbound}, the flow can follow two different routes depending on the stability of the current-free equilibrium \eqref{eqn:current-free}. Although a linear stability analysis is outside the scope of this paper, we can identify the cause of the instability in our simulations. 

In ideal MHD, weakly magnetized Keplerian flows are subject to the MRI \citep{balbushawley91}. In a 1D shearing box (vertical structures only), the MRI can be stabilized by both Ohmic \citep{jin96} and ambipolar diffusion \citep{desch04,kunzbalbus04}. The Hall effect can be either stabilizing or destabilizing depending on the strength and orientation of the net magnetic field \citep{wardle99,balbusterquem01}. For the Hall-dominated regime probed in this paper, the HSI appears while the MRI is resistively damped \citep{kunz08,wardlesalmeron12}. 

If the current-free equilibrium is linearly stable, then the MHD diffusivities dissipate electric currents and let the flow relax to the same equilibrium. Otherwise, the initial perturbations grow exponentially in time and amplify the electric current and magnetic stress through the disk. Since the 1D shearing box forbids the development of `parasitic' secondary instabilities \citep{goodmanxu94, latter10, kunzlesur13}, the exponential growth saturates in the non-linear regimes of Hall and ambipolar diffusion. This saturation happens before magnetic pressure significantly alters the disk structure. 

The linear phase of the instability is illustrated in Appendix \ref{app:integrator}. The dissipative effects allow the system to reach a steady-state which only depends on the choice of $(r,\beta)$ and not on the initial noise. We qualify these states as `internally-driven' because they are powered by the orbital shear and satisfy \eqref{eqn:alphadisk}. For the magnetizations $\beta\in\left[10^3, 10^8\right]$ considered, the $B_z<0$ cases are always linearly stable; we therefore focus on the $B_z>0$ cases in this section. 

\subsection{Reference solution} \label{sec:refint}

We start by exhibiting the properties of a reference solution computed at $r=2\au$ with $\beta=10^7$. Since the relative importance of non-ideal MHD effects varies with radius, we provide a second example solution computed at $r=1\au$ in Appendix \ref{app:example2}.

\subsubsection{Vertical structure}

\begin{figure}
\begin{center}
\includegraphics[width=\columnwidth]{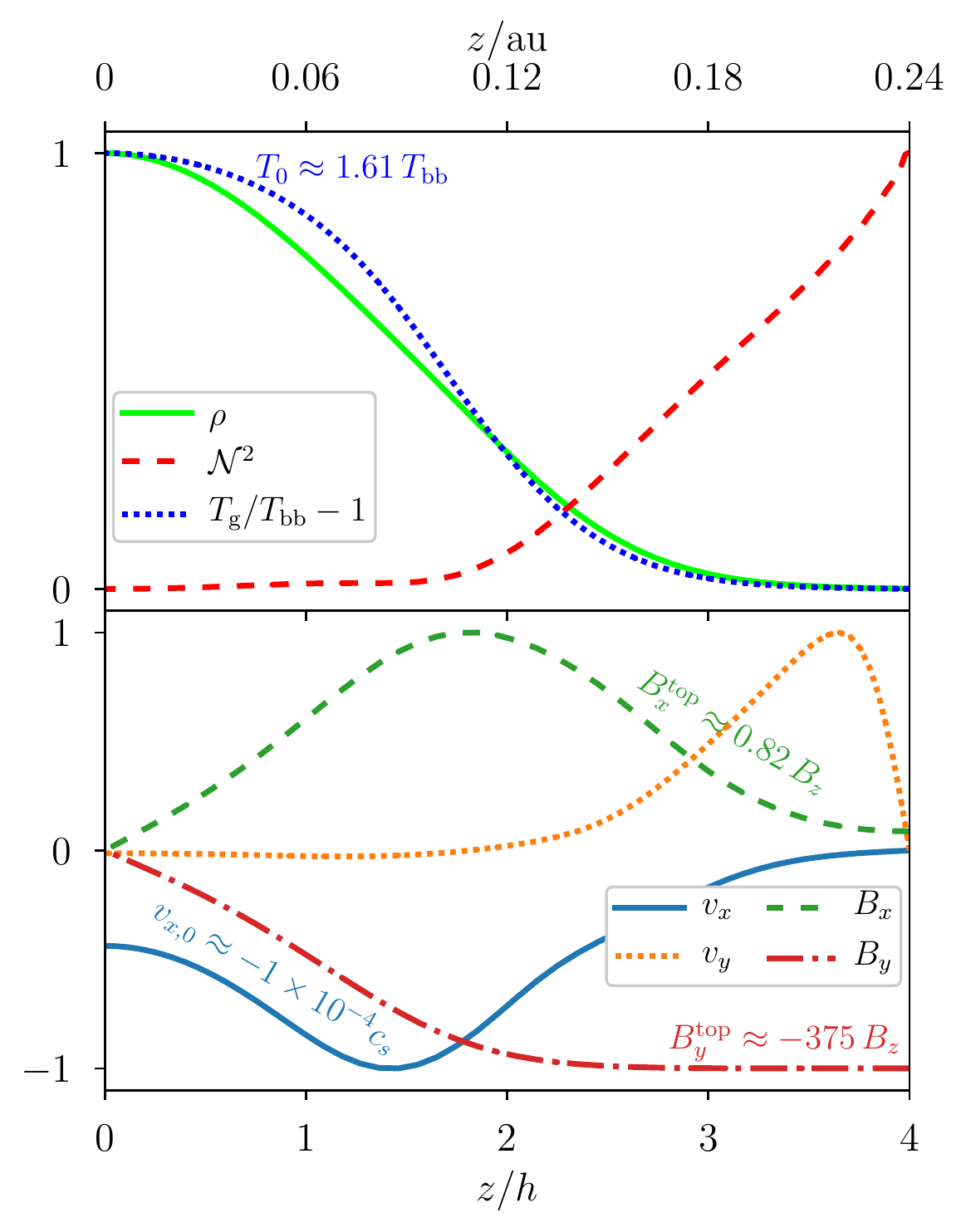}
\caption{Vertical profiles of the flow variables in an equilibrium with $\beta=10^7$ at $r=2\au$, normalized by their extremal value for visibility. \emph{Upper panel:} density (solid green), squared \BV frequency (dashed red) and gas temperature relative to $\Tbb$ (dotted blue). \emph{Lower panel:} radial velocity (solid blue), azimuthal velocity (dotted orange), radial magnetic field (dashed green), azimuthal magnetic field (dot-dashed red). \label{fig:mripr1k1b7_fulleq}}
\end{center}
\end{figure}

Fig. \ref{fig:mripr1k1b7_fulleq} shows the vertical profiles of the flow variables in a steady state computed at $r=2\au$ with $\beta=10^7$; the curves have been rescaled to fit in $\left[-1,1\right]$ for visibility.

The thermodynamic variables are represented on the upper panel. The density distribution is close to Gaussian; it is always decreasing with height ($\partial_z\rho \leq 0$), as for every equilibria presented in this paper. The gas temperature is maximal in the midplane where it reaches $\Tg \approx 1.61 \Tbb$. It is equal to the radiation temperature to $10^{-6}$ accuracy on the entire interval (not shown). The squared \BV frequency $\mathcal{N}^2>0$ everywhere, so this equilibrium is convectively stable. However, $\mathcal{N}^2 \approx 0$ for $z/h\lesssim 2$ implies that the deep disk is close to marginal stability. Other solutions do exhibit entropy profiles decreasing with height, see Sect. \ref{sec:convection}. 

The MHD variables are represented on the lower panel, where the velocities correspond to deviations from the Keplerian background. The azimuthal velocity $v_y$ is negative near the midplane and positive in the surface layers, indicating that angular momentum has been exchanged between the two layers. The radial velocity $v_x$ has a constant sign, so there is a net mass accretion rate $\dot{M} = \int \rho v_x \dd z \neq 0$ in the entire domain\footnote{The system \eqref{eqn:dtrho}-\eqref{eqn:dter} of the shearing-sheet equations is independent of $x$, so $\int \rho v_x \dd z\neq 0$ can be interpreted as mass accretion regardless of its sign.}. The midplane radial velocity is only $-1\times 10^{-4} c_s$ so the accretion flow is very sub-sonic. The horizontal magnetic field ($B_x,B_y$) grows from zero in the midplane to its maximal amplitude over a scale $\sim h$. The product $-B_x B_y \geq 0$ generates a radial flux of angular momentum (Maxwell stress) through the disk. The azimuthal component $\Byt \approx -375 B_z$ at the upper boundary, so the magnetic field is tightly coiled and the magnetic pressure $B^2/2 \gtrsim \rho c_s^2$ above $2h$.

The relation \eqref{eqn:stressacc} is satisfied by construction: the $-\Byt B_z$ stress removes angular momentum vertically and causes an accretion rate $\dot{M}\neq 0$ even in the absence of an outflow ($v_z=0$). Unlike in global disk models, the radial flux of angular momentum --- measured by $\alpha\approx 7\times 10^{-4}$ in \eqref{eqn:alphadisk} for this equilibrium --- cannot cause a net mass accretion rate in the shearing box. 

\subsubsection{Energy budget}

\begin{figure}
\begin{center}
\includegraphics[width=\columnwidth]{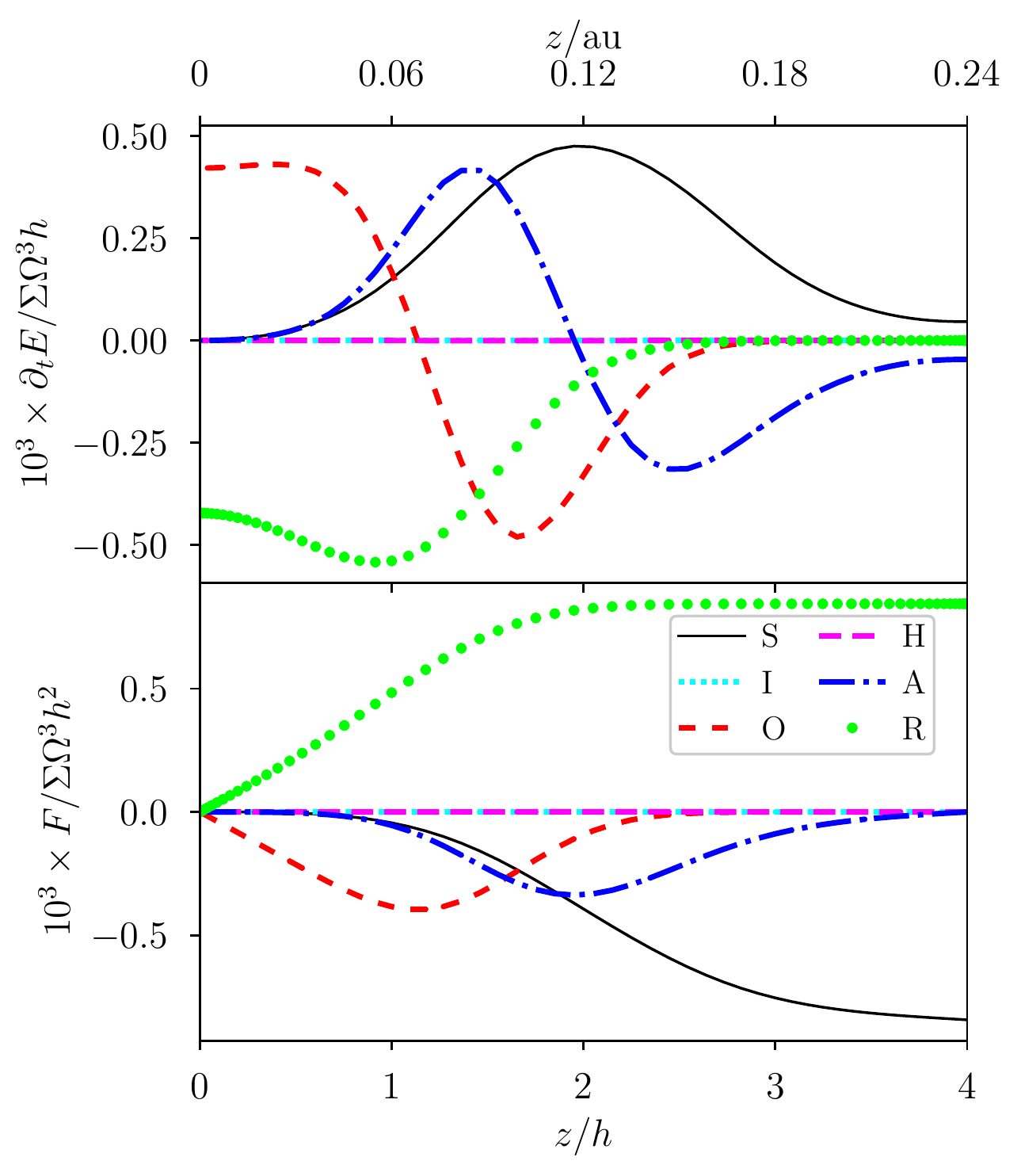}
\caption{Energy budget in the same equilibrium as on Fig. \ref{fig:mripr1k1b7_fulleq}. \emph{Upper panel:} total energy equation \eqref{eqn:nrj}. \emph{Lower panel:} associated energy fluxes, equivalent to the vertical integral of \eqref{eqn:nrj} from the midplane. The different curves correspond to the source term (`S', solid black), the ideal induction (`I', dotted cyan), Ohmic resistivity (`O', dashed red), the Hall drift (`H', dashed magenta), ambipolar diffusion (`A', dot-dashed blue) and radiation (`R', green dots). \label{fig:mripr1k1b7_fullnrj}}
\end{center}
\end{figure}

To explain the buildup of heat in the midplane, we decompose the evolution of the total energy density in the reference simulation and plot the result in Fig. \ref{fig:mripr1k1b7_fullnrj}. The individual terms of \eqref{eqn:nrj} are represented on the upper panel. The associated fluxes are drawn on the lower panel, where the source term (`S') is integrated vertically from the midplane and multiplied with a minus sign to allow comparison with the other fluxes. 

On the upper panel, the source term (solid black) represents the extraction of energy from the Keplerian shear into velocity and magnetic fields by the $xy$ Reynolds and Maxwell stresses. It is maximal near $z\approx 2h$ and positive at every altitudes, increasing the total energy relative to the current-free equilibrium. 

The Ohmic (dashed red) and ambipolar (dot-dashed blue) terms both have two distinct effects on the energy content of the plasma. On one side, they locally dissipate magnetic energy into thermal energy, with no effect on the total energy density. On the other side, they diffusively spread magnetic energy away from its maximum, causing the downward energy fluxes drawn on the lower panel of Fig. \ref{fig:mripr1k1b7_fullnrj}. These fluxes vanish at the boundaries of the computational domain, so they induce no net energy gain nor loss in the equilibrium. Ambipolar diffusion dominates in the upper layers $z/h\gtrsim 2$. Ohmic diffusion is predominant at low altitudes $z/h \lesssim 1$ and it is the only term bringing energy down to the midplane. The Hall term (dashed magenta) can only transport energy via waves. The associated energy flux is negligible in this equilibrium. 

On the upper panel, the radiative term (green dots) is negative everywhere so it removes energy from the equilibrium. This is achieved by an upward radiative flux $F\lR \geq 0$ on the lower panel, transporting radiative energy from the midplane out of the disk. The radiative flux increases with height and becomes roughly constant above $z/h \gtrsim 2$, so the conversion from thermal to radiative energy happens mostly below this height. It is the only term balancing the net energy input caused by the source term and allowing the system to reach a thermodynamic equilibrium. 

\begin{figure}
\begin{center}
\includegraphics[width=\columnwidth]{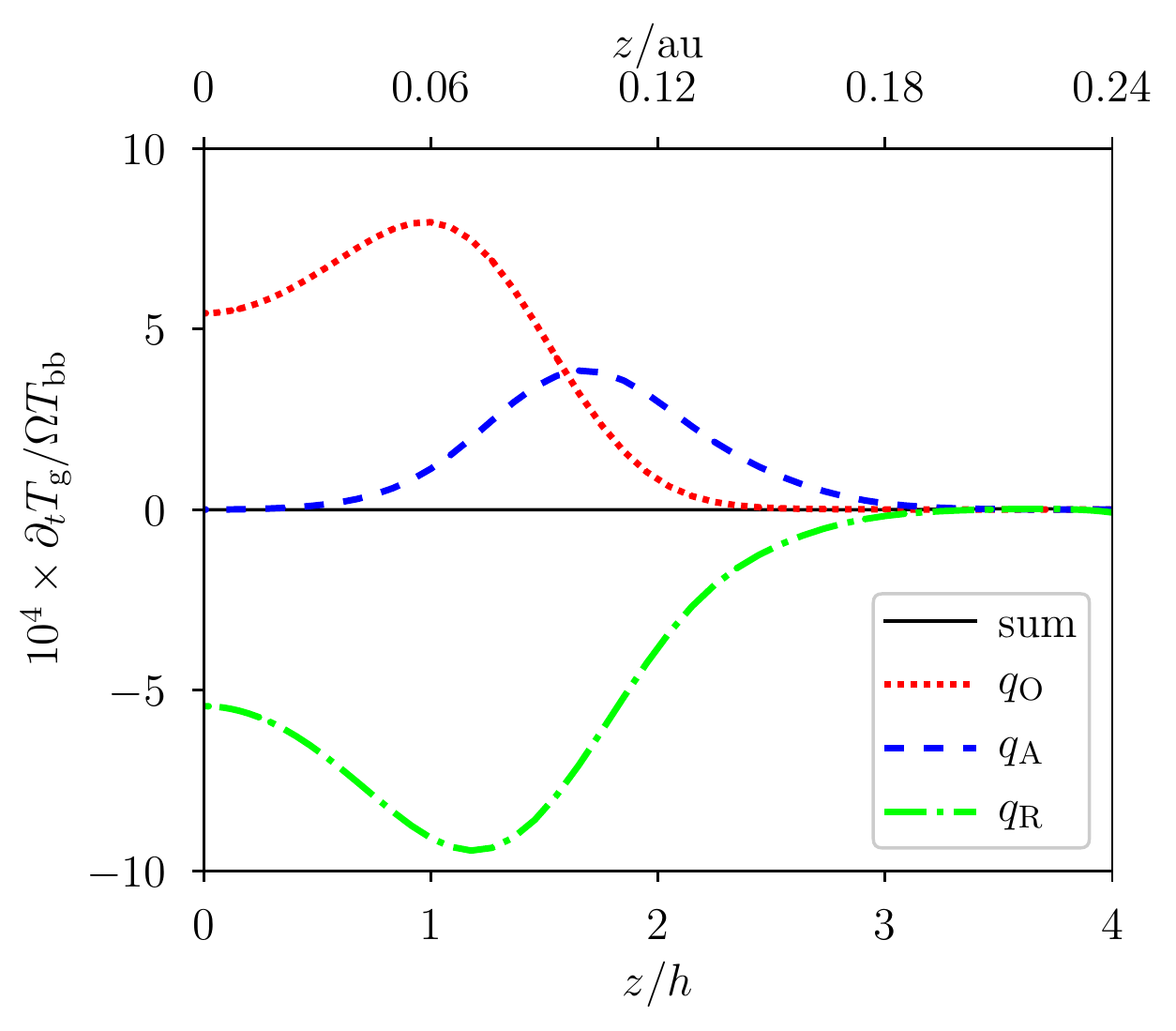}
\caption{Heating/cooling rates per unit mass as defined in \eqref{eqn:qheat} for the same equilibrium as on Fig. \ref{fig:mripr1k1b7_fulleq}. Ohmic ($q\lO$, \emph{dotted red}) and ambipolar heating ($q\lA$, \emph{dashed blue}) are balanced by radiative cooling ($q\lR$, \emph{dot-dashed green}). \label{fig:mripr1k1b7_heat}}
\end{center}
\end{figure}

Examining the total energy budget does not reveal where the conversion from kinetic and magnetic to internal energy happens. To clarify which effect is responsible for heating the gas, we decompose the internal energy (pressure) equation into specific heating/cooling rates as in \eqref{eqn:qheat} on Fig. \ref{fig:mripr1k1b7_heat}. The Ohmic and ambipolar resistivities both dissipate magnetic energy (electric currents) into heat. Electric heating is localized near the midplane and is primarily caused by Ohmic resistivity for this specific equilibrium. The radiative cooling rate $q\lR$ is the same as on the upper panel of Fig. \ref{fig:mripr1k1b7_fullnrj} after dividing by $\rho$, as expected from \eqref{eqn:dtp}-\eqref{eqn:dter} in steady state. 

A key feature of these equilibria is that the energy extracted from the shear is redistributed vertically before being thermalized. Heating can thus occur in the midplane although the stress extracts orbital energy away from the midplane. This feature is more prominent on the alternative example solution provided in Appendix \ref{app:example2}. 

\subsection{Dependence on disk magnetization}

We proceed to explore how our main diagnostics depend on the plasma $\beta \in \left[10^4,10^8\right]$ at four different radii $r/\au\in\left\lbrace 0.5,1,2,4\right\rbrace$ from the star. We consistently find a one to one mapping between $\left(r,\beta\right)$ and the steady state profiles. Hence, at a given radius there seems to exist a single solution branch parametrized by $\beta$ and stable to z-dependent perturbations. The remaining parameters are kept the same as in the reference equilibrium described above.

As a precaution, we stopped the exploration when the temperature contrast $T_0/\Ttau-1$ reached unity. For larger temperature constrasts the disk becomes geometrically thicker, so our numerical domain may become insufficiently large to describe the solutions adequately. The effective scale-height of the disk, measured as the standard deviation of a gaussian profile fitting the density distribution below $3h$, is always less than $1.5h$.

\subsubsection{Angular momentum transport and associated heating} \label{sec:mriheat}

\begin{figure}
\begin{center}
\includegraphics[width=\columnwidth]{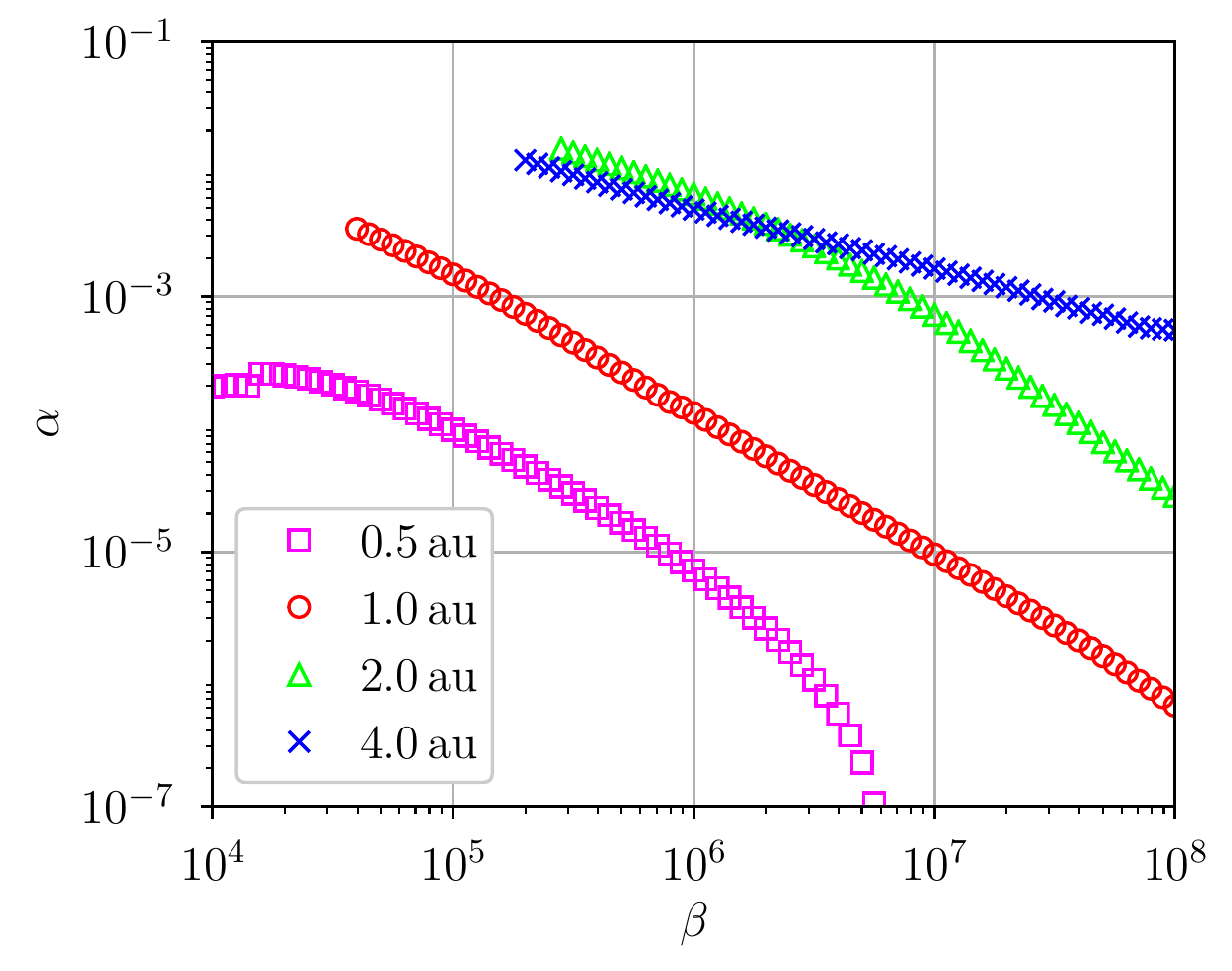}
\caption{Dissipation coefficient $\alpha$ as a function of the plasma $\beta$ at different radii from the star (see legend). We stopped the exploration when the temperature contrast reached unity. \label{fig:mripr1k1_alpha}}
\end{center}
\end{figure}

We start by quantifying the radial flux of angular momentum and the associated resistive heating as expressed by \eqref{eqn:alphadisk}. For each equilibrium we measure the coefficient $\alpha$ as defined by \eqref{eqn:defalpha} and place it on Fig. \ref{fig:mripr1k1_alpha}. Since we exclude solutions with a temperature contrast larger than unity, there are solutions at larger magnetizations (smaller $\beta$) than we show here: the breaks in our solutions branches are just where we end our parameter scan. 

The dissipation coefficient $\alpha$ is a decreasing functions of $\beta$ at the four radii considered. At $1\au$ we find that $\alpha$ scales roughly as $\beta^{-1/2}$, i.e., increases as $B_z$, and the scaling becomes shallower at larger radii. The values of $\alpha$ range from $10^{-7}$ to $10^{-2}$ over this parameter space. For a given $\beta$, the coefficient $\alpha$ increases by more than one order of magnitude from $0.5\au$ to $1\au$ and by another order of magnitude from $1\au$ to $2\au$. We stopped the exploration when the temperature contrast in the disk reached unity, but more solutions presumably exist with $\alpha \gtrsim 10^{-2}$ for lower $\beta$. These levels of laminar magnetic stress are in agreement with the 3D stratified shearing box simulations of \citet{LKF14} which included the Hall effect. Ohmic heating dominates over ambipolar heating by less than a factor $10$ at $0.5$ and $1\au$, they become comparable at $2\au$ and ambipolar heating dominates at $4\au$ (not shown). 

\subsubsection{Temperature contrast}

\begin{figure}
\begin{center}
\includegraphics[width=\columnwidth]{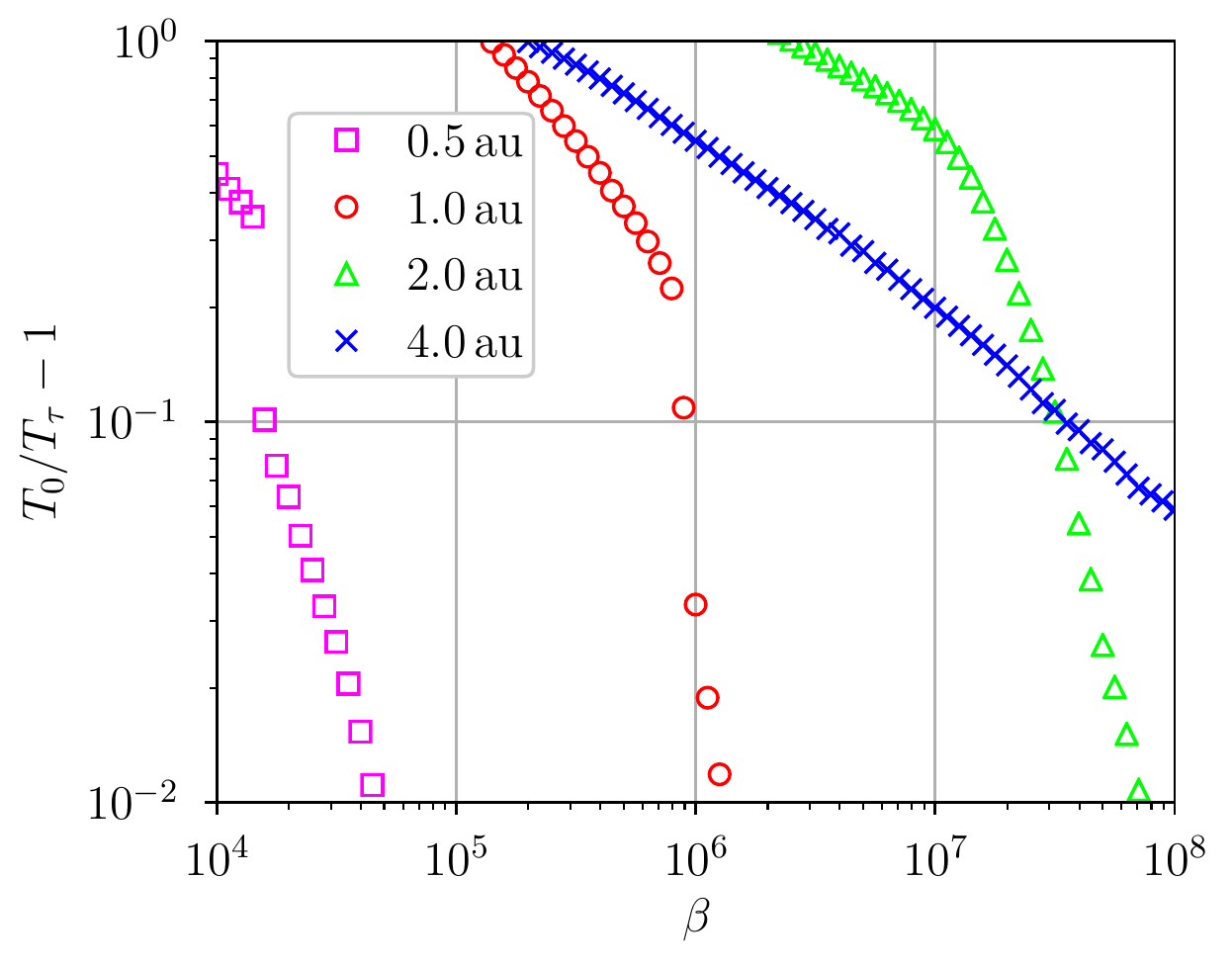}
\caption{Temperature contrast between the midplane and the $\tau=1$ altitude as a function of the plasma $\beta$ at different radii from the star (see legend). \label{fig:mripr1k1_tmp}}
\end{center}
\end{figure}

Following on from the heating efficiency of these solutions, we evaluate the temperature contrast achieved between the midplane and the $\tau=1$ altitude. Fig. \ref{fig:mripr1k1_tmp} represents this temperature contrast measured for each pair of parameters $\left(r,\beta\right)$.

The temperature contrast is a decreasing function of $\beta$ at any given radius, so the more magnetized the disk the hotter the midplane compared to the surface. The temperature contrast reaches unity for $\beta \in \left[10^5, 3\times 10^6\right]$ at $r\geq 1\au$. When going from small to larger radii, the temperature contrast becomes flatter as a function of $\beta$. At $r=1\au$ the temperature contrast decreases from $1$ to $10^{-2}$ over a single decade of $\beta \in \left[1\times 10^5, 1\times 10^6\right]$. At $r=2\au$ the temperature contrast is larger than $10^{-2}$ over $\beta\in\left[3\times 10^6,7\times 10^7\right]$. At $r=4\au$ the temperature contrast is already larger than $10^{-2}$ at the largest $\beta=10^8$ considered. Order unity differences between the midplane and surface temperatures are therefore achievable at all radii for sufficient disk magnetization.

\subsubsection{Convective stability} \label{sec:convection}

For each equilibrium represented on Fig. \ref{fig:mripr1k1_tmp} we compute the entropy profile via \eqref{eqn:bvn2} and deduce the profile of the squared \BV frequency $\mathcal{N}^2$. If $\mathcal{N}^2<0$ over a range of altitudes, then this range could be convectively unstable if we permitted perturbations with a radial dependence. The characteristic timescale for the growth of convective modes would then be $\omega \equiv \sqrt{-\mathcal{N}^2}$. Because a full linear stability analysis is outside the scope of this paper, we focus on this necessary condition for convection. 

\begin{figure}
\begin{center}
\includegraphics[width=\columnwidth]{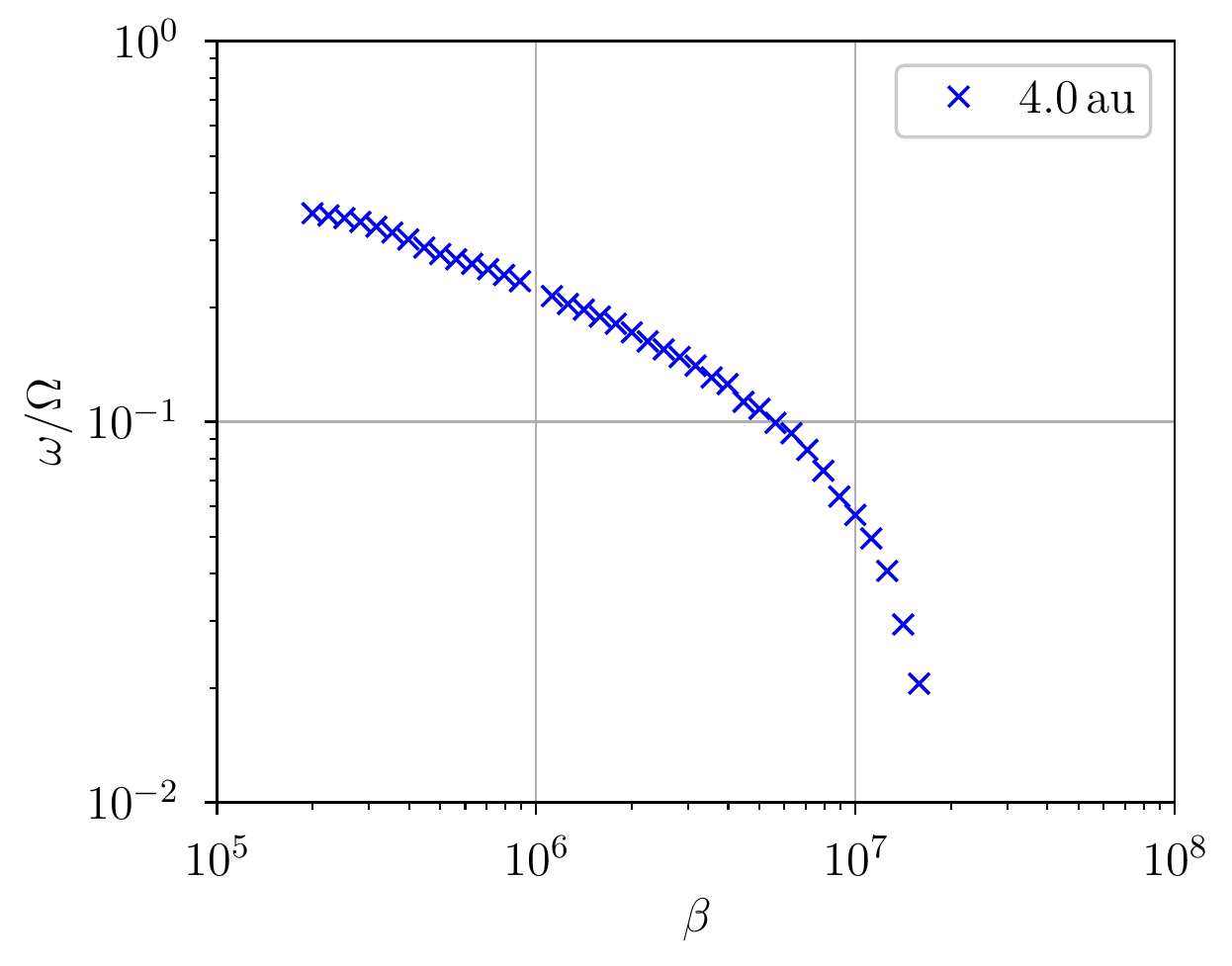}
\caption{Maximal \BV growth rate $\omega/\Omega$ as a function of the plasma $\beta$, computed from the entropy profile of equilibria at different radii from the star (see legend). We stopped the exploration when the temperature contrast reached unity. \label{fig:mripr1k1_bvx}}
\end{center}
\end{figure}

We show the maximal value of $\omega/\Omega$ measured in each equilibria over this parameter space on Fig. \ref{fig:mripr1k1_bvx}. In this disk model, only the equilibria at $r=4\au$ have a reversed entropy gradient leading to $\omega>0$ over a range of altitudes. At $r=4\au$ the temperature contrast reaches unity for $\beta\approx 3\times 10^5$; more solutions with $\omega>0$ presumably exist at lower $\beta$, which we excluded by precaution. 

The \BV growth rates range from a few $10^{-2}$ to over $3 \times 10^{-1} \Omega$ for the equilibria represented on Fig. \ref{fig:mripr1k1_bvx}. The range of altitudes over which $\omega>0$ spans roughly $z/h \in \left] \,0,1\right]$ and it expands to higher altitudes from the midplane as $\beta$ decreases. Given the absence of viscosity and the slow radiative timescale following from the assumed opacity, these equilibria could support unstable convective motions. 

None of the solutions computed at $r\leq 2\au$ have a reversed entropy gradient despite reaching order-unity temperature contrasts. Upon inspection of these solutions, the temperature profiles are flatter near the midplane, see for example Fig. \ref{fig:mripr1k1b6_fulleq} in Appendix \ref{app:example2}. When the gas temperature starts decreasing with $z$, the density is decreasing faster so that the specific entropy $s = \Tg / \rho^{\gamma-1}$ remains monotonically increasing. To understand what controls the temperature gradient, we rewrite the radiative term in \eqref{eqn:qheat} assuming a steady state radiation field in \eqref{eqn:dter}:
\begin{equation}
  \rho q\lR = -\left(\gamma-1\right) \nabla\cdot F\lR.
\end{equation}
In the optically thick regions $F\lR \simeq -\left(\lambda c/3\right)\nabla E\lR$ with $E\lR\simeq a\Tg^4$ to first order. The steady-state gas temperature then satisfies
\begin{equation} \label{eqn:conductor}
  \nabla\cdot\left(\lambda \nabla\Tg^4\right) + \frac{3}{\left(\gamma-1\right) a c} \rho \left(q\lO+q\lA\right)\simeq 0,
\end{equation}
so the midplane behaves as a thermal conductor with Ohmic and ambipolar heating acting as source terms. Flat temperature profiles $\partial_z \Tg \ll \Tbb/\lambda$ therefore occur when electric heating is localized away from the midplane. At $r=0.5\au$ and $1\au$ electric heating is indeed less efficient below $z\lesssim 2h$ (see Fig. \ref{fig:mripr1k1b6_fullnrj}). 

\subsection{Dependence on ionization fraction} \label{sec:lowion}

\begin{figure}
\begin{center}
\includegraphics[width=\columnwidth]{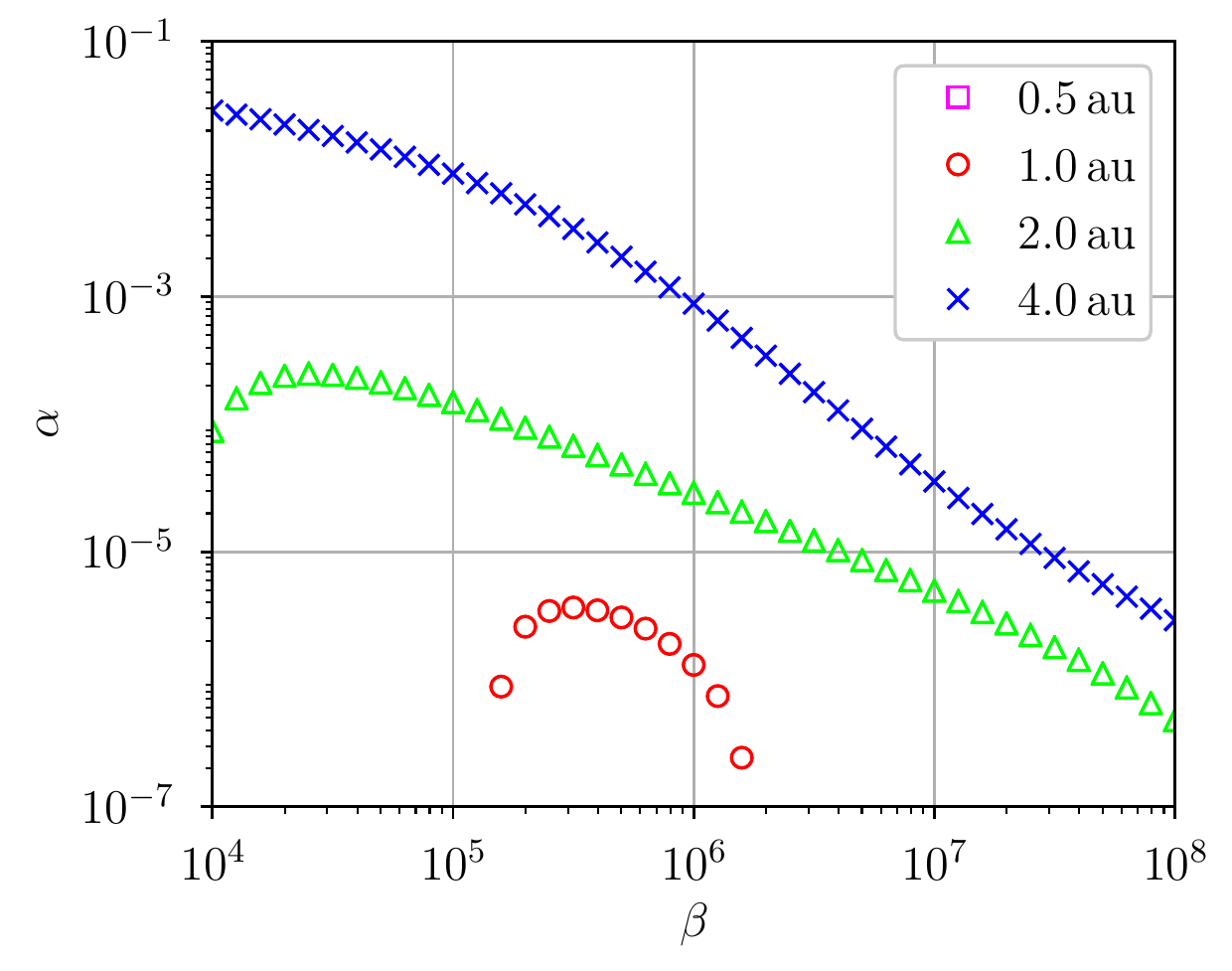}
\caption{Same as Fig. \ref{fig:mripr1k1_alpha} with an ionization fraction $x_e$ reduced by a constant factor $10^{-2}$ at every height $z/h$. \label{fig:mripr4k1_alpha}}
\end{center}
\end{figure}

To account for the possible influence of dust grains on molecular recombination and charge capture, we reduce the ionization fraction $x_e$ by a constant factor $10^{-2}$ at every height. This is equivalent to increasing the MHD diffusivities $\eta \propto 1/x_e$. We show on Fig. \ref{fig:mripr4k1_alpha} how the dissipation coefficient $\alpha$ varies with $r$ and $\beta$ in this more resistive case compared to Sect. \ref{sec:mriheat}. 

At $r=0.5\au$ the disk is linearly stable over the entire range of $\beta$ considered: the disk converges to the current-free equilibrium \eqref{eqn:current-free} and therefore supports no electric heating. At $r=1\au$ the disk is linearly unstable only for $\beta \gtrsim 10^5$. The resulting $\alpha$ are maximal near $\beta \approx 3\times 10^5$ and remain weaker than $\alpha \lesssim 10^{-5}$. At $r=2\au$ the dissipation coefficient reaches $\alpha\approx 2\times 10^{-4}$ for $\beta=2\times 10^4$. At $r=4\au$ the disk is linearly unstable for the whole range of $\beta$ considered and $\alpha$ reaches roughly $3\times 10^{-2}$ for $\beta = 10^4$. In comparison with Fig. \ref{fig:mripr1k1_alpha}, the dissipation coefficients are $10^{-1}$ to $10^{-3}$ times lower for a given $\beta$.

\begin{figure}
\begin{center}
\includegraphics[width=\columnwidth]{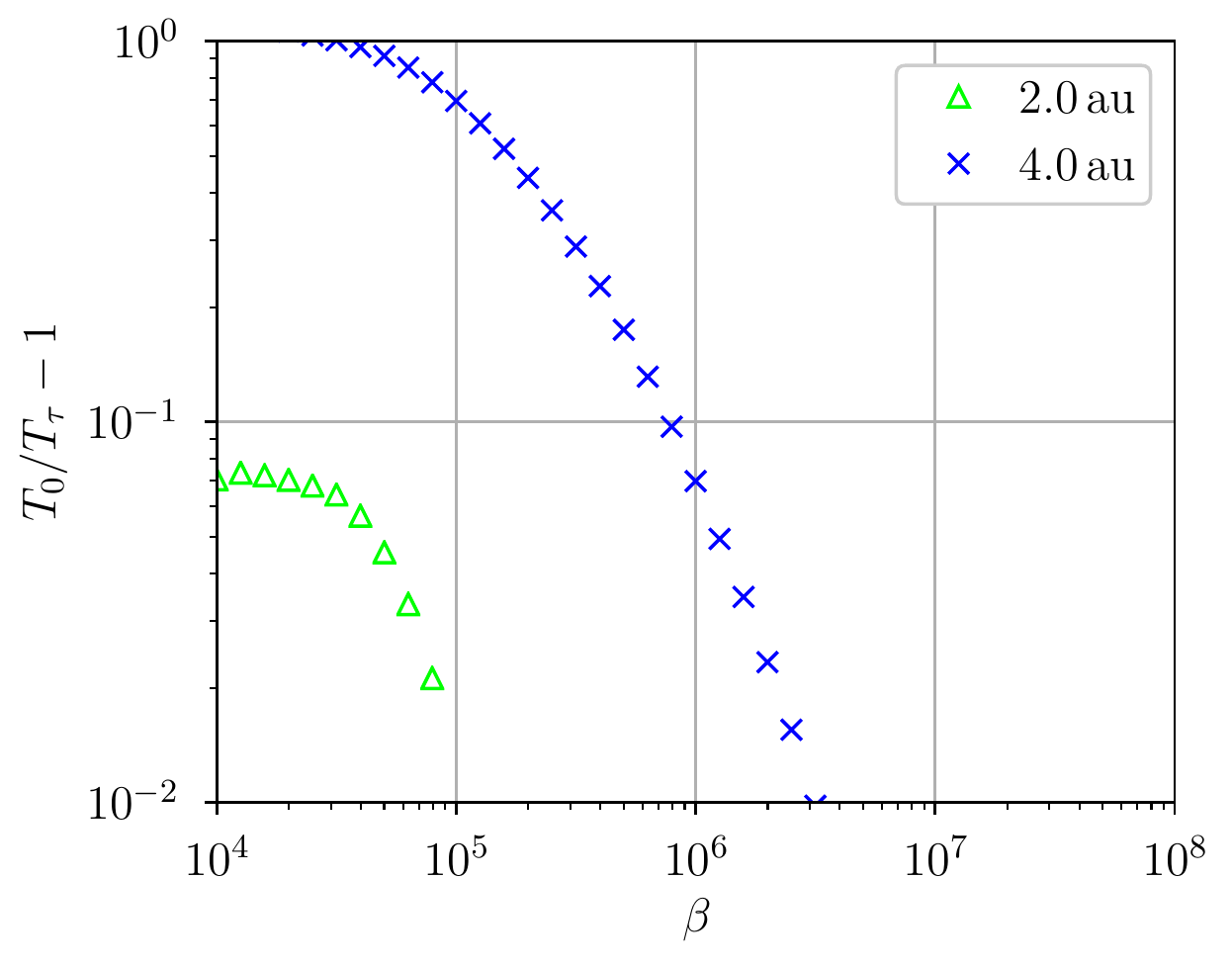}
\caption{Same as Fig. \ref{fig:mripr1k1_tmp} with an ionization fraction $x_e$ reduced by a constant factor $10^{-2}$; the temperature contrast remains below $10^{-2}$ at $r=0.5\au$ and $1\au$. \label{fig:mripr4k1_tmp}}
\end{center}
\end{figure}

Fig. \ref{fig:mripr4k1_tmp} shows the temperature contrast $T_0/\Ttau-1$ measured in the same series of equilibria as on Fig. \ref{fig:mripr4k1_alpha}. At $r=0.5\au$ and $1\au$ the temperature contrast remains below $10^{-2}$ for this range of disk magnetizations. At $r=2$ and $4\au$ the temperature contrast reaches its maximal value $7\times 10^{-2}$ and $1$ respectively for $\beta\approx 2\times 10^4$. The ionization fraction thus controls to a large degree the efficiency of electric heating and angular momentum transport inside the disk.

\subsection{Dependence on gas opacity} \label{sec:opacity}

For the range of temperatures considered, the frequency-integrated opacity is dominated by dust grains \citep{belllin94,ferguson05}. In the absence of vertical mixing, the sedimentation of dust grains toward the disk midplane and their coagulation would lower the average opacity. To account for the possible sedimentation and coagulation of dust grains, we reduce the opacity by a constant factor $10^{-1}$ with respect to the values of \citet{belllin94}.

At $1\au$, the altitude of the $\tau=1$ surface in the current-free equilibrium \eqref{eqn:current-free} becomes $z_{\tau}/h\approx 2.99$, so most of the gas in the computational domain is still confined to high optical depths. The $\tau=1$ altitude decreases to $z_{\tau} \approx 2.50h$ at $r=2\au$ and to $z_{\tau} \approx 1.86h$ at $r=4\au$. In this last case, most of the computational domain is transparent to radiation so the vertical FLD approximation becomes inappropriate. We therefore exclude the $r=4\au$ case in this section. 

\begin{figure}
\begin{center}
\includegraphics[width=\columnwidth]{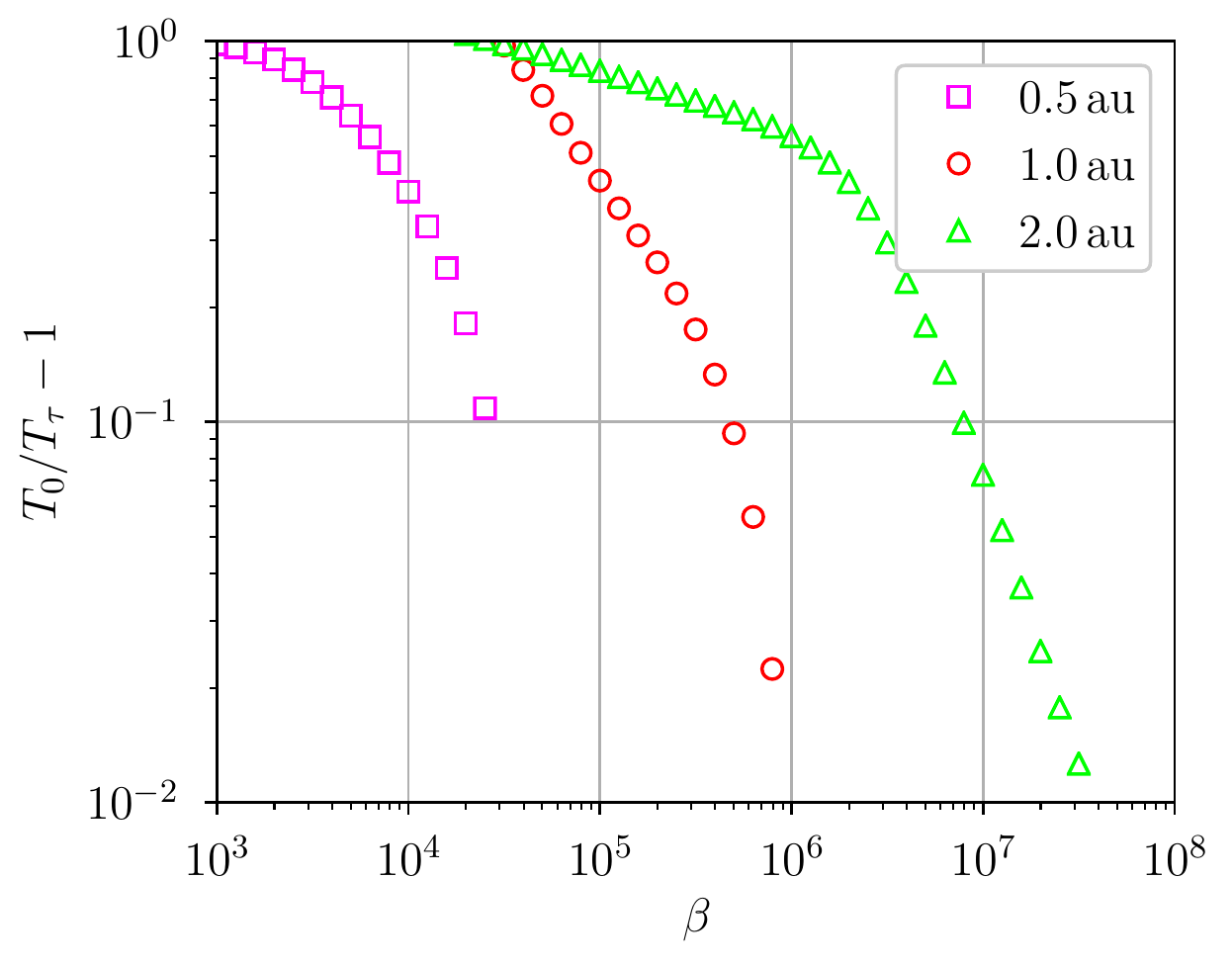}
\caption{Same as Fig. \ref{fig:mripr1k1_tmp} with a reduced opacity $\kappa/10$. \label{fig:mripr1k01_tmp}}
\end{center}
\end{figure}

Fig. \ref{fig:mripr1k01_tmp} shows the temperature contrast obtained as a function of $r$ and $\beta$ when the opacity is reduced by a factor $10^{-1}$ relative to our reference case. This figure is qualitatively similar to Fig. \ref{fig:mripr1k1_tmp}, but the temperature contrast at a given $\beta$ is now reduced by a factor $\approx 1/10$. At $r=2\au$, when $\beta$ decreases from $10^6$ to $2\times 10^4$ the temperature contrast increases slowly from $50$ per cent to unity, although the heating rate $\alpha$ increases from $10^{-2}$ to $10^{-1}$ (not shown). This suggests that an increasing fraction of the heating power is injected in the optically thin layers and immediately radiated away. A stronger magnetization is thus required to reach the same temperature contrast in low-opacity disks.

\subsection{Midplane symmetry in two-sided disks} \label{sec:oddsymm}

In the shearing box simulations of \citet{LKF14}, \citet{bai2015} and \citet{mori+19}, the flow spontaneously adopts an `odd' symmetry about the midplane. Instead of the equatorial symmetry \eqref{eqn:lowbound} which we imposed, the odd symmetry is such that
\begin{equation} \label{eqn:symm}
  \left(v_x,v_y,B_x,B_y\right)\left(-z\right) = \left(-v_x,-v_y,+B_x,+B_y\right)\left(z\right).
\end{equation}
This symmetry allows no electric current in the midplane, so electric heating is necessarily localized in the surface layers $\vert z/h\vert \gtrsim 2$ \citep{mori+19}. Since the Maxwell stress $-B_y B_z$ has a constant sign, angular momentum is injected from one boundary and extracted from the other, causing no net accretion through the disk. This odd symmetry is not only an artifact of the shearing box as it was also obtained in global disk simulations \citep{gressel15,bethune17,suriano18,rodenkirch20}. 

We tested the stability of the reference solution of Sect. \ref{sec:refint} in the two-sided domain $z/h \in \left[-4,+4\right]$. We used the symmetrized one-sided solution as initial condition, perturbed it and repeated the time integration until reaching a steady state. The flow converged to the solution with an equatorial symmetry $\left(B_x,B_y\right)\left(-z\right)=\left(-B_x,-B_y\right)\left(z\right)$ matching our standard midplane conditions \eqref{eqn:lowbound}. The midplane symmetry considered in this paper is therefore 1D stable for at least a range of radii and magnetizations. 

\section{Externally-driven states} \label{sec:ext}

\subsection{Rationale} \label{sec:extstates}

In Sect. \ref{sec:lowion} we reduced the ionization fraction $x_e$ by a factor $10^{-2}$ and the current-free equilibrium \eqref{eqn:current-free} became linearly stable at $r=0.5\au$ regardless of $\beta$. When decreasing the ionization fraction by a factor $10^{-3}$, the current-free equilibrium is in fact stable for all $\beta$ and radii up to $r=4\au$ in our disk model. Although the instability is quenched in this regime, electric currents may still flow through the disk if an external magnetic torque $-\Byt B_z$ acts on the disk surface. In this case, the energy dissipated inside the disk is not extracted from the orbital shear but instead provided at the surface by an energy flux $F_z^{\mathrm{top}}$ as in \eqref{eqn:alphatorque}. 

A variety of situations can lead to such external torques if the disk is threaded by a large-scale poloidal field. These include magnetized winds \citep{pudritznorman83,pelletierpudritz92} regardless of their launching mechanism --- photoevaporative or magnetocentrifugal. Since the key to mass accretion and energy dissipation is the magnetic stress $-B_y^{\mathrm{top}} B_z$, as a first approximation we can neglect the outflow velocity $v_z$ in the energetic balance \citep[see for example][]{lovelace02, lovelace09}. Alternatively, the magnetic field threading the disk might be anchored in a well ionized medium with a different rotation rate (e.g., the star, the infalling cloud, or a different radius of the disk). 

In this section we repeat the previous calculations with two major changes. First, we decrease the ionization fraction by a factor $10^{-3}$ with respect to Sect. \ref{sec:ionfraction} so that the current-free equilibrium \eqref{eqn:current-free} is linearly stable at every radius and magnetizations considered. Second, we impose the value of $\Byt \equiv B_y(4h)$ at the surface of the disk, which sets the flux of angular momentum leaving the disk $-\Byt B_z$ and the resulting mass accretion rate via \eqref{eqn:stressacc}. By fixing $\Byt$, we set the magnetic energy flux $F_z^\mathrm{top}$ entering the disk in \eqref{eqn:alphatorque}. To keep the number of free parameters to a minimum, we impose $\beta=10^5$ and only vary the radius $r$ and surface azimuthal field $\Byt$. 

\subsection{Reference solution} \label{sec:refext}

\subsubsection{Vertical structure} \label{sec:refextvert}

\begin{figure}
\begin{center}
\includegraphics[width=\columnwidth]{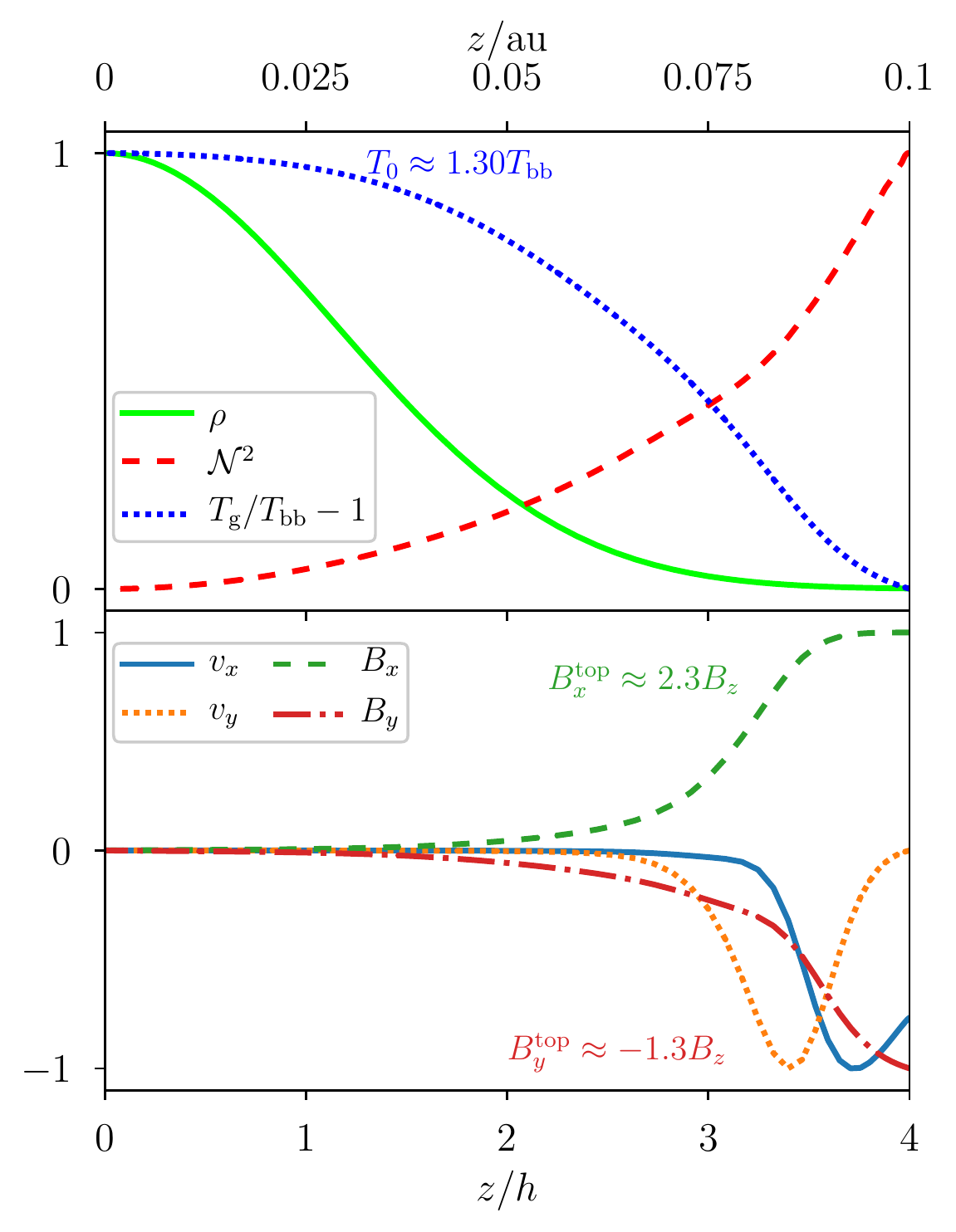}
\caption{Vertical profiles of the flow variables in an externally-driven equilibrium at $r=1\au$ with a surface azimuthal field $\Byt=-2\times 10^{-2} \,\mathrm{G}$. The curves are normalized by their extremal value for visibility. \emph{Upper panel:} density (solid green), squared \BV frequency (dashed red) and gas temperature relative to $\Tbb$ (dotted blue). \emph{Lower panel:} radial velocity (solid blue), azimuthal velocity (dotted orange), radial magnetic field (dashed green), azimuthal magnetic field (dot-dashed red). \label{fig:windpr6k1b5by002G_fulleq}}
\end{center}
\end{figure}

Fig. \ref{fig:windpr6k1b5by002G_fulleq} shows the vertical profiles of the flow variables in an externally-driven equilibrium at $r=1\au$ with $\beta = 10^5$ and a surface azimuthal field $\Byt=-2\times 10^{-2} \,\mathrm{G} \approx -1.26 B_z$. The curves have been rescaled to fit in $\left[-1,1\right]$ for visibility. 

As in the internally-driven case of Fig. \ref{fig:mripr1k1b7_fulleq}, the upper panel shows that the gas density is nearly Gaussian and the squared \BV frequency $\mathcal{N}^2>0$ at every altitude. The gas and radiation temperatures are equal to better than $10^{-6}$ accuracy everywhere (not shown) and maximal in the midplane with $T_0 \approx 1.30\Tbb$.

The lower panel shows that the velocity and magnetic perturbations are localized above $z/h\gtrsim 2$. The radial velocity $v_x<0$ in an accretion layer around $z/h \approx 3.5$. Similarly the azimuthal velocity $v_y<0$ in a narrow layer centered on $z/h\approx 3.4$. $B_y\approx 0$ in the midplane and its growth toward $\Byt$ happens rapidly above $z\gtrsim 3h$. The radial component $B_x$ must only satisfy $\partial_z B_x=0$ at the top boundary, so its amplitude $\Bxt \approx 2.3 B_z$ is not fixed a priori. In this case $B^2/2 \lesssim P$ at the surface of the disk, i.e. the magnetic field is nearing equipartition as expected for general accretion-ejection structures \citep{ferreira93}. From the surface toward the midplane, $B_x$ decays over a characteristic length scale $\sim h$. Despite the electric currents being localized near the disk surface, we still find a significant build up of heat in the midplane. 

\subsubsection{Energy budget} \label{sec:extnrj}

\begin{figure}
\begin{center}
\includegraphics[width=\columnwidth]{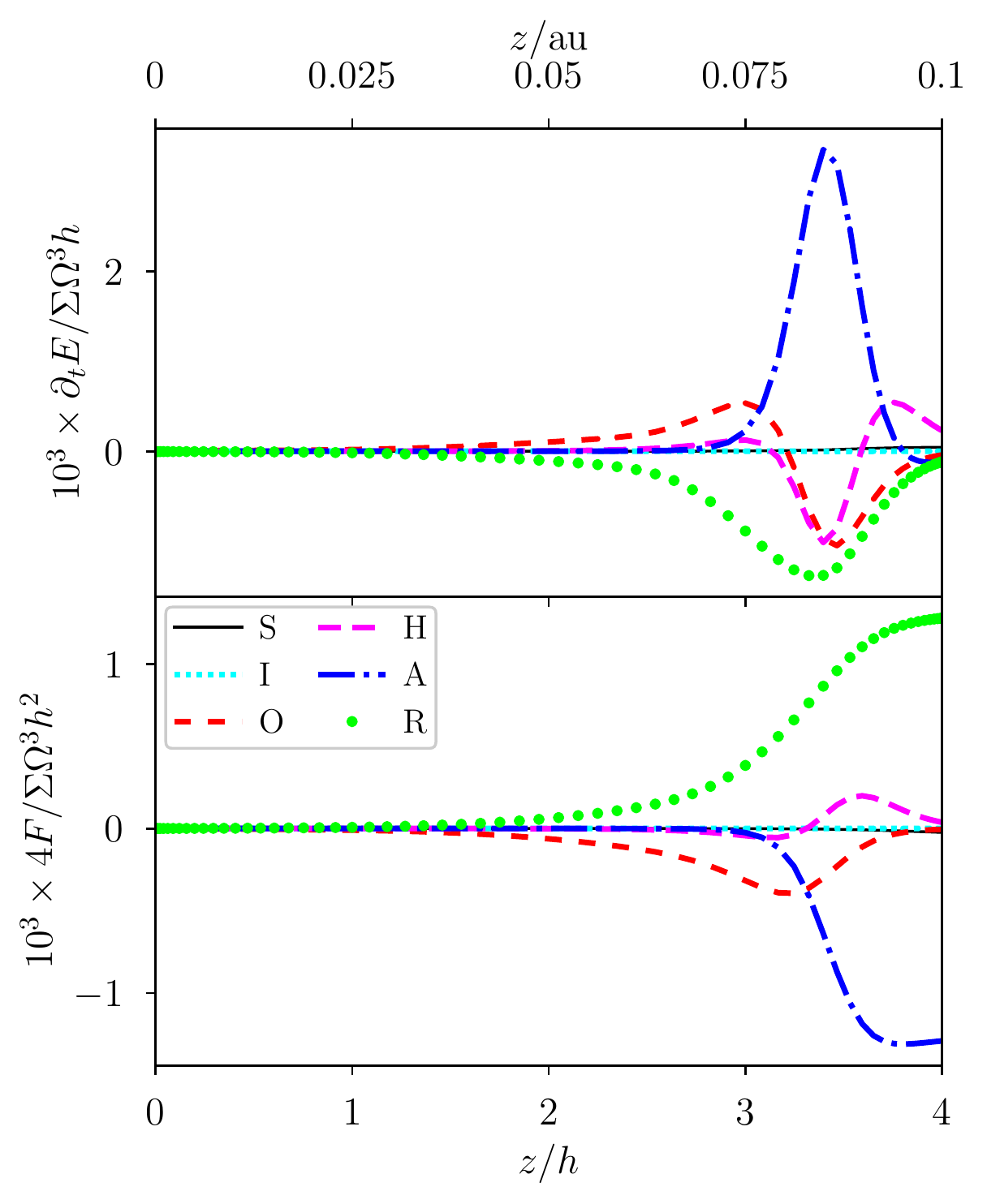}
\caption{Same as Fig. \ref{fig:mripr1k1b7_fullnrj} in the externally-driven equilibrium of Fig. \ref{fig:mripr1k1b7_fulleq}. The source term 'S' (solid black) is nearly zero everywhere, so energy is mainly supplied at the upper boundary. \label{fig:windpr6k1b5by002G_fullnrj}}
\end{center}
\end{figure}

On Fig. \ref{fig:windpr6k1b5by002G_fullnrj} we disentangle the energy exchanges in the previous externally-driven equilibrium. A key difference with the internally-driven state shown on Fig. \ref{fig:mripr1k1b7_fullnrj} is that the internal stress / source term (solid black) is now negligible compared to every other contributions by two orders of magnitude. Although the product $-B_x B_y$ is non-zero on Fig. \ref{fig:windpr6k1b5by002G_fulleq}, the resulting stress extracts a negligible amount of energy from the Keplerian shear. In contrast to the internally-driven solution \eqref{fig:mripr1k1b7_fullnrj}, the energy input in this equilibrium is mainly supplied at the upper boundary and satisfies the balance \eqref{eqn:alphatorque}.

As previously, the Ohmic and ambipolar fluxes transport energy downward from the disk surface. At the upper boundary, it is ambipolar diffusion that brings the energy of the imposed $\Byt$ (and resulting $\Bxt$) from the surface $z=4h$ into the disk. The Ohmic term becomes dominant near $z/h \approx 3$ and extends down to the midplane. The radiative flux is oriented upward in the entire domain, evacuating heat through the upper boundary and allowing the system to reach a thermodynamic equilibrium. 

The vertically-integrated Ohmic and ambipolar heat fluxes amount to a dissipation coefficient $\alpha \approx 7\times 10^{-5}$ in this equilibrium. As shown above and in \eqref{eqn:alphatorque}, this $\alpha$ is not related to the vertically-integrated stress but to the dissipation of the energy supplied at the disk surface. We can use \eqref{eqn:conductor} to interpret the build up of heat in the midplane despite electric heating being localized near $z/h \approx 3$. Since the photon mean free path increases with height ($\partial_z\lambda>0$), radiative diffusion favors temperature maxima in the midplane ($\partial_z^2 \Tg<0$) as long as it is optically thick\footnote{During the evolution toward a steady state, the temperature initially rises where heat is deposited ($z/h\approx 3$) but this is only a transient stage.}. 

\subsection{Dependence on surface magnetic field} \label{sec:windheat}

\begin{figure}
\begin{center}
\includegraphics[width=\columnwidth]{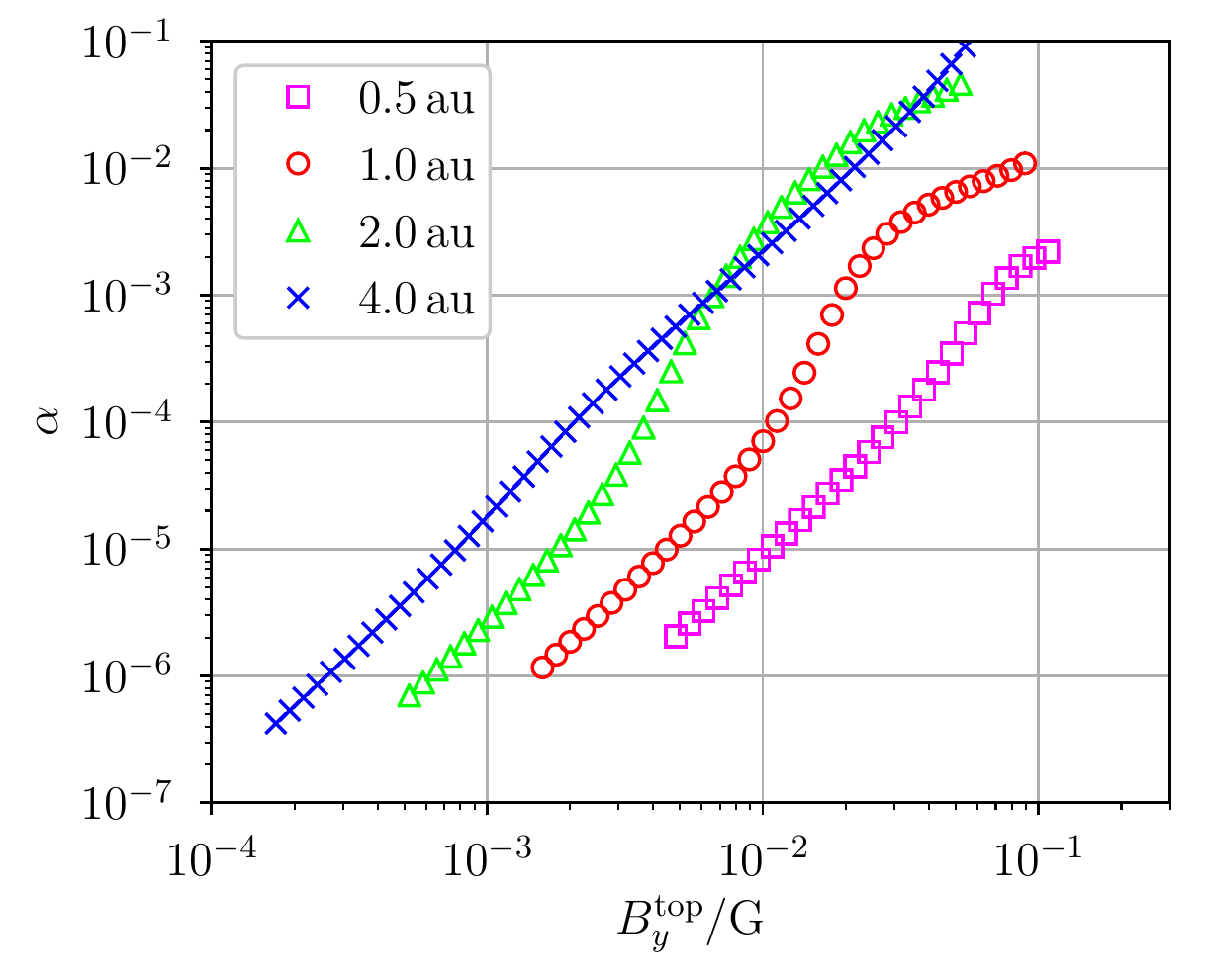}
\caption{Dissipation coefficient $\alpha$ as a function of the surface azimuthal field $\Byt$ at different radii from the star (see legend) in externally-driven equilibria with $\beta=10^5$. Each curve starts at $\Byt/B_z=10^{-1}$. \label{fig:windp_alpha}}
\end{center}
\end{figure}

We now sample different radii $r$ and azimuthal magnetic fields $\Byt$ while keeping every other parameter fixed as in Sect. \ref{sec:refext}. For each externally-driven solution we compute the dimensionless dissipation coefficient $\alpha$ via \eqref{eqn:defalpha} and represent it on Fig. \ref{fig:windp_alpha}. The magnetic field $\Byt$ is arbitrarily measured in Gauss units.

The dissipation coefficient $\alpha$ increases with $\Byt$ at the four radii considered. Its values range from $\alpha=10^{-7}$ to $10^{-1}$ for the interval of $\Byt / \,\mathrm{G} \in \left[ 10^{-4}, 10^{-1} \right]$ considered. As mentioned in Sect. \ref{sec:refextvert}, the surface magnetic field is near equipartition with the gas pressure for $\Byt/\,\mathrm{G}=2\times 10^{-2}$ at $r=1\au$. All the solutions presented on Fig. \ref{fig:windp_alpha} are in this regime of moderate magnetization. 

We find a roughly quadratic scaling of the dissipation coefficient $\alpha$ with $\Byt$, most apparent at low $\Byt$ or in the $r=4\au$ case. We also note that the graphs of the dimensional heat flux $Q\lO+Q\lA$ as a function of $\Byt$ (not shown) are nearly superimposed on each other, so the separation between the four solution branches on Fig. \ref{fig:windp_alpha} mainly reflects variations of the normalization coefficient $\Sigma h^2 \Omega^3$ with radius. Since the energy dissipated in the disk is related to the surface energy flux $F_z^{\mathrm{top}}$ via \eqref{eqn:alphatorque}, it does not have to scale with the characteristic energy flux of the disk $\Sigma h^2 \Omega^3$. 

\begin{figure}
\begin{center}
\includegraphics[width=\columnwidth]{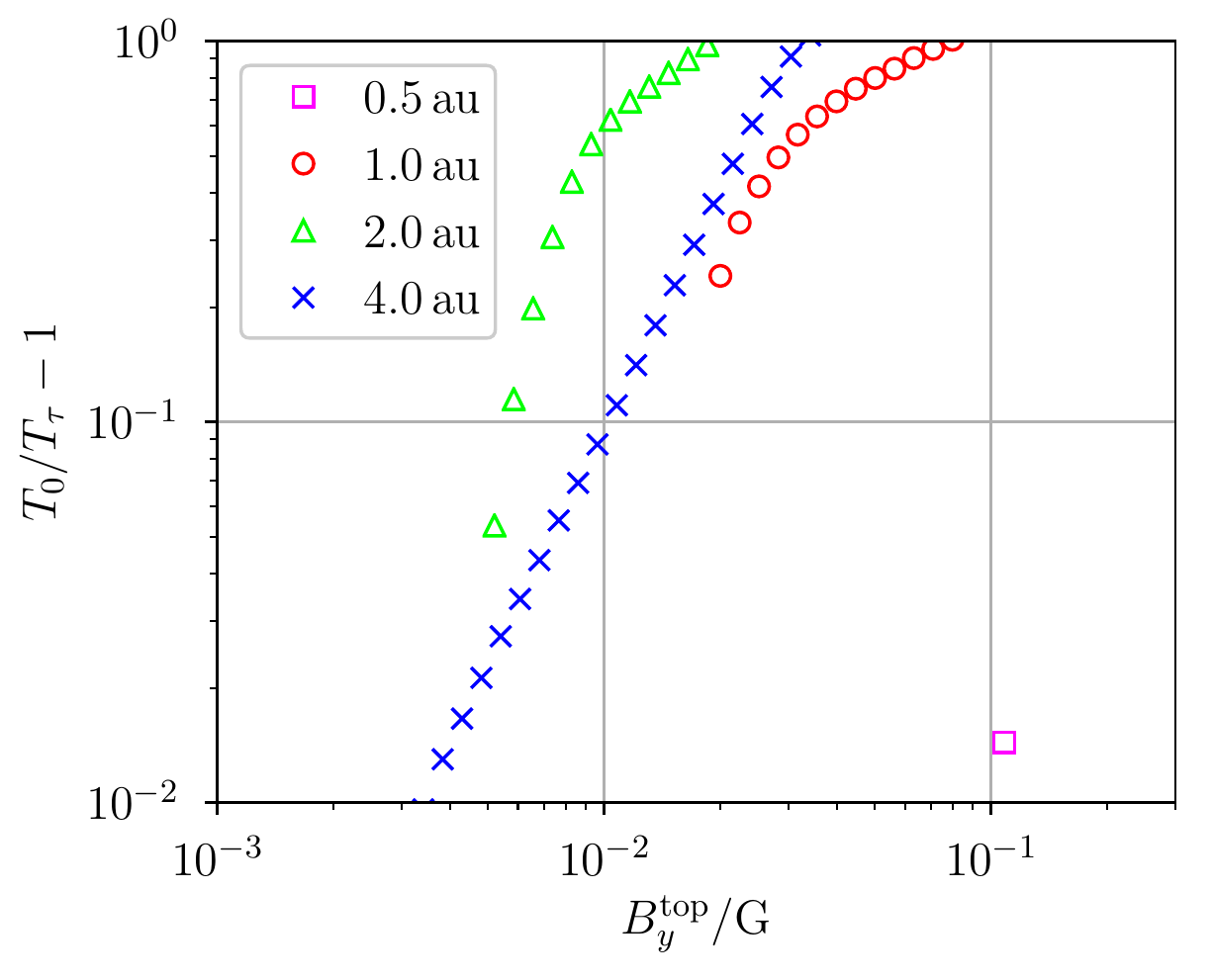}
\caption{Temperature contrast between the midplane and the $\tau=1$ altitude as a function of the surface azimuthal field $\Byt$ at different radii from the star (see legend) in externally-driven equilibria with $\beta=10^5$. \label{fig:windp_tmp}}
\end{center}
\end{figure}

On Fig. \ref{fig:windp_tmp} we show the temperature contrast $T_0/\Ttau-1$ measured in the same series of equilibria as on Fig. \ref{fig:windp_alpha}. At $r\geq 1\au$ the temperature contrast increases from $10^{-2}$ to $\sim 1$ over one decade in $\Byt$. This range of $\Byt$ is independent of the strength of the vertical field $B_z$ threading the disk as long as $\beta\gg 1$ because only $\left(B_x,B_y\right)$ can generate electric currents in this 1D model, and the net $B_z$ only weakly contributes to the ambipolar diffusivity. 

When considering the opposite polarity $B_z<0$ for the net magnetic field, we found that the current-free equilibrium was always linearly stable and therefore supported no energy dissipation by itself. This polarity dependence is introduced by the Hall effect in the induction equation. Since the externally-driven states do not rely on a linear instability as the energy source, we can freely impose an energy flux $F_z^{\mathrm{top}}$ on a disk with $B_z<0$ and obtain different dissipation properties than in the $B_z>0$ case. 

\begin{figure}
\begin{center}
\includegraphics[width=\columnwidth]{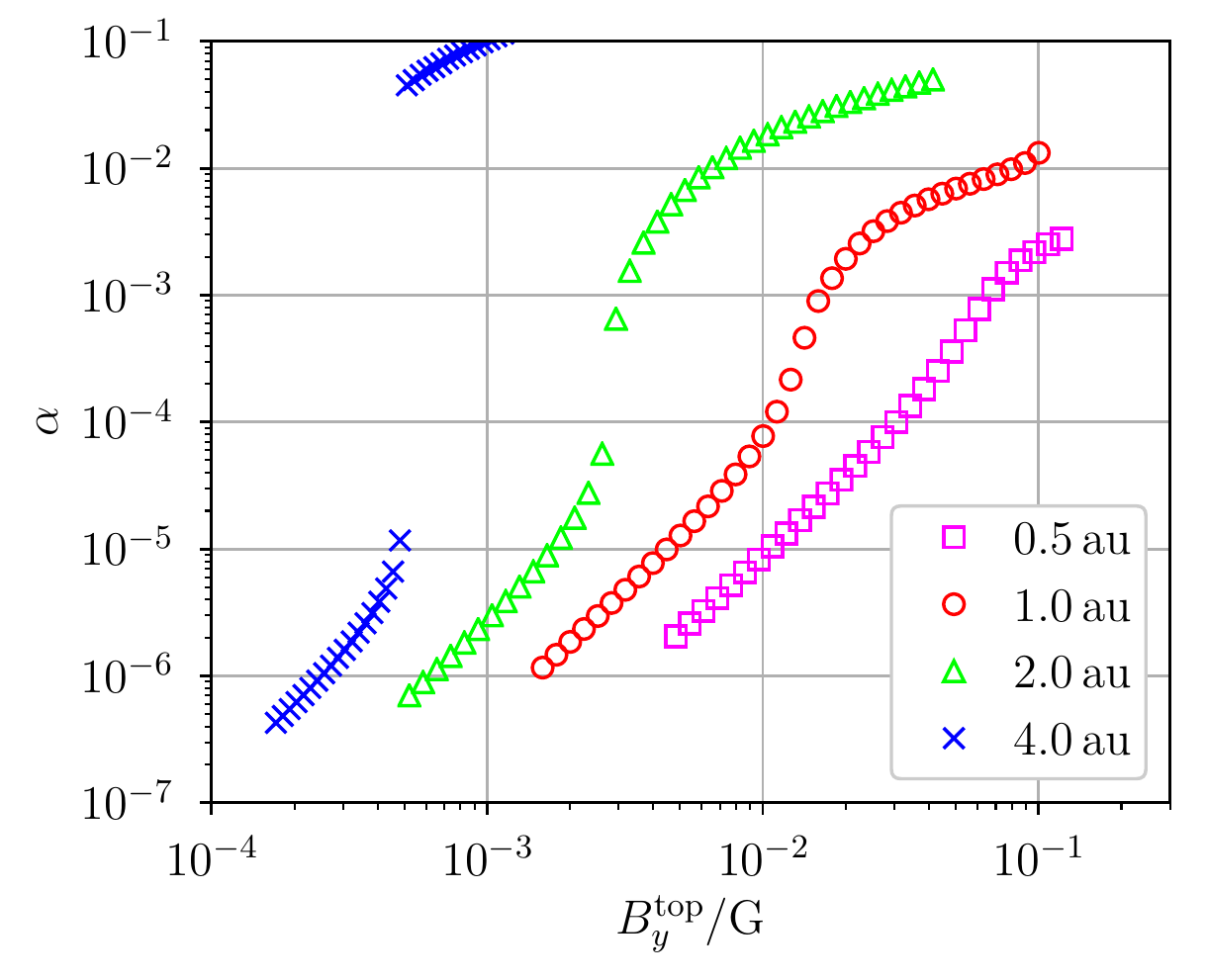}
\caption{Same as Fig. \ref{fig:windp_alpha} with a reversed polarity for the net magnetic field $B_z<0$. \label{fig:windm_alpha}}
\end{center}
\end{figure}

We show on Fig. \ref{fig:windm_alpha} the dissipation coefficient $\alpha$ obtained when repeating the same parameter sampling as for Fig. \ref{fig:windp_alpha} but with the opposite polarity for the net magnetic field: $B_z<0$. As previously, $\alpha$ is an increasing function of $\Byt$ at every radii and spans the range $\left[10^{-7},10^{-1}\right]$ over the interval of $\Byt$ considered.

The solution branches corresponding to $r=0.5\au$ and $1\au$ are qualitatively the same as on Fig. \ref{fig:windp_alpha}. At $r=1\au$ the dissipation coefficient $\alpha$ is greater than in the $B_z>0$ case by at most a factor $\approx 4$. The difference becomes more significant at $r=2\au$: the dissipation coefficent reaches $\alpha=10^{-2}$ for a surface $\Byt \approx 7\times 10^{-3}\,\mathrm{G}$, three times smaller than in the $B_z>0$ case. At $r=4\au$ the solution branch already deviates from the $B_z>0$ case for $\Byt \gtrsim 4\times 10^{-4}\,\mathrm{G}$, and $\alpha$ becomes over a thousand times larger than in the $B_z>0$ case for surface fields as weak as $\Byt \lesssim 6\times 10^{-3}\,\mathrm{G}$. 

Interestingly, the $B_z<0$ orientation of the net magnetic field leads to larger dissipation rates for a given magnetic field $\Byt$ at the surface of the disk. Since the energy dissipated in the disk is equal to the energy input at the top boundary, it implies that the energy flux $F_z^{\mathrm{top}}$ is larger in norm when $B_z<0$. Because only the Hall effect induces such a polarity dependence, we can attribute the enhanced $F_z^{\mathrm{top}}$ to the contribution of the Hall energy flux at the surface of the disk. Reciprocally, it takes a larger energy flux at the surface of the disk to maintain a given $\Byt$ when $B_z<0$. 

\section{Discussion} \label{sec:discussion}

\subsection{Caveats} \label{sec:caveats}

\subsubsection{Chemistry and radiation}

We neglected the influence of dust grains and metals in the ionization model used throughout this paper. The midplane abundance of free electrons can decrease by several orders of magnitude depending on the dust grain properties alone \citep{ivlev16}. Moreover, dust grains can directly alter the electric resistivity when considered as dominant charge carriers themselves \citep{salmeron08,ilgner12,xubai16}. Acknowledging these uncertainties, we opted for a plausible ionization model and covered the `dead-zone' regime when artificially decreasing the ionization fraction by a factor $10^{-2}$ in Sect. \ref{sec:lowion} and $10^{-3}$ in Sect. \ref{sec:ext}.

Inversely, the prescribed opacity relies on $\sim 20\,\mu\mathrm{m}$ dust grains at the temperatures considered $\sim 150\,\mathrm{K}$. This prescription should be valid in the inner few $\au$ while the disk is optically thick to its own thermal radiations \citep{chiangoldreich97,alessio01}. We obtained qualitatively similar results when reducing the opacity by a factor ten in Sect. \ref{sec:opacity}, but larger magnetizations were then required to heat the midplane to a given temperature.

Our model does not capture the deposition of heat by stellar photons at the disk surface and its reprocessing to thermal wavelengths. By solving for the reprocessed radiation only, we exclude the temperature stratification separating the disk from the warm stellar environment \citep{aresu11,pinte+18}. Solving for the deposition of heat as implemented by \citet{flock13} and \citet{kolb13} should yield the same temperature in the midplane as long as the disk is thick to both the incomming and reprocessed radiation. As mentioned in Sect. \ref{sec:ionfraction}, the disk surface should also be better ionized above $z/h\gtrsim 4$ by stellar FUV photons \citep{perezchiang11} which we omitted.

\subsubsection{Imposed symmetries}

The assumption of axisymmetry is supported in good approximation by both local and global disk simulations including Ohmic and ambipolar diffusion in the regime relevant to protoplanetary disks \citep{baistone11,gressel15}.

For simplicity we have imposed the symmetry \eqref{eqn:lowbound} of the disk about the midplane. The symmetry supported by global disk simulations incorporating all three non-ideal MHD effects can differ from \eqref{eqn:lowbound} or keep evolving with radius and time \citep{bethune17}. In Sect. \ref{sec:oddsymm} we relaxed the midplane conditions and verified that the flow is attracted to the same equatorial symmetry for a specific set of input parameters. However, we cannot exclude that other solutions would rather bifurcate to an `odd' midplane symmetry if permitted as in \citet{LKF14} and \citet{mori+19}. 

Our most questionable assumption is that of radial locality, and its implications are twofold. On one hand, we found that electric heating and angular momentum transport strongly depend on the conditions inside the disk. Placing our local simulations back in a global disk picture, the measure of stress $\alpha$ could vary on $\au$ scales and induce radially inhomogeneous accretion. The resulting gas build-ups could in turn have implication on the migration of solids \citep{weidenschilling77} or the generation of vortices \citep{lovelace99,meheut12}. The non-ideal MHD effects also affect the radial transport of magnetic flux \citep{guilet13,guilet14,leung19}. Our model does not permit this redistribution of mass and magnetic flux across neighboring annuli. 

On the other hand, this 1D model excludes radially dependent waves and instabilities. The entropy profiles examined in Sect. \ref{sec:convection} suggest that some equilibria might be convectively unstable in more than one dimension \citep{ruden88,kley93,held18}. The strong azimuthal magnetic field could trigger the growth of oblique modes of the ambipolar-shear instability \citep{kunzbalbus04,kunz08}. In principle, the azimuthal velocity profiles $\partial_z v_y \ll \Omega$ could also trigger the vertical shear instability \citep{urpin98}, albeit with small growth rates and saturation amplitudes in the optically thick regions considered \citep{nelson13,linyoudin15}. Our solutions might therefore be unstable and evolve toward new, not necessarily stationnary states in more than one dimension.

Finally, our boundary conditions prevent the removal of energy by advection ($v_z=0$) or by the ideal MHD Poynting flux associated with torsional Alfv\'en waves. This Poynting flux could be used to power outflows in jet-launching disks, so the heating rates obtained in our model should be regarded as upper limits given the orbital energy extracted by the magnetic field \citep{casse2000}.

\subsection{Laminar versus turbulent disks}

Assuming that mass accretion occurs via laminar magnetic torques in the inner few $\au$, our model produces a wide range of $\alpha \in \left[10^{-7},10^{-1}\right]$ depending on the ionization and magnetization of the disk. Values as high as $\alpha \gtrsim 10^{-2}$ would not necessarily induce measurable accretion rates since they are restricted to narrow radial ranges, favoring the formation of substructures in the disk instead.

Large values $\alpha > 10^{-4}$ are noteworthy for several reasons. First, they support the possibility of magnetically-driven accretion in regions where the MRI should be resistively damped \citep{gammie96}. The Hall effect plays a major role in destabilizing the disk \citep{balbusterquem01,wardlesalmeron12}, and in producing a laminar magnetic stress as obtained in both local \citep{LKF14,bai2015} and global simulations \citep{bethune17,bai2017}. On the other hand, should the resistivity be large enough to stabilize the disk (e.g., at lower ionization fractions), electric currents originating from large-scale accretion-ejection structures may still permeate the disk \citep{ferreira97}, inducing significant heating and accretion inside $z/h \lesssim 3$.

Second, turbulent motions inferred from molecular line broadening and dust lifting seem compatible with $\alpha_{\mathrm{SS}} \sim 10^{-4}$ \citep{flaherty15,flaherty17,pinte16}, where $\alpha_{\mathrm{SS}}$ is the \citet{shakusunyaev73} measure of turbulent stress. If angular momentum was only transported by turbulence, then this $\alpha_{\mathrm{SS}}$ should match the one measured by turbulent heating and mass accretion \citep{balbuspap99}. However, if the magnetic stress is primarily laminar then the heating and mass accretion rates can be much larger than those suggested by the gas kinematics alone. While the magnetic field provides most of the stress in a laminar fashion, hydrodynamic or dust-gas instabilities may be responsible for turbulence at the observed amplitudes $\alpha_{\mathrm{SS}} \sim 10^{-4}$ \citep[e.g.,][]{youdin05,fromang19}.

\subsection{Comparison with the work of Mori et al. 2019}

The closest study to ours is that of \citet{mori+19}, who concluded that electric heating should have a negligible impact on the temperature of the inner disk compared to stellar irradiation. There are a number of differences between their model and ours which hinder a direct quantitative comparisons. In particular:
\begin{enumerate}
\item they deduce the thermal structure of the disk from the MHD heating rates computed in an isothermal disk. Although this method is not self-consistent, we believe that it can be a good approximation as long as the obtained temperature contrast is smaller than unity.
\item they consider gas opacities ranging from $5\times 10^{-1}$ to $5\times 10^{-3}\,\mathrm{cm}^2\,\mathrm{g}^{-1}$, i.e., ten to one thousand times lower than our reference case. Their largest opacity is comparable to our reduced opacity case of Sect. \ref{sec:opacity}, where we obtained consequently lower temperature contrasts. We believe that this is the main cause of discrepancy between our results and those of \citet{mori+19}. Our results support their conclusion of inefficient accretion heating in the limit of low disk opacities. 
\item They let the disk evolve in the two-sided domain $z/h \in \left[-8,+8\right]$ and only obtain the `odd' midplane symmetry discussed in Sect. \ref{sec:oddsymm}. This symmetry supports no electric current in the disk midplane, whence electric heating is localized away from the midplane and does not favor heat accumulation inside the disk.
\end{enumerate}

The method of \citet{mori+19} might also introduce biases in the energy budget which are difficult to quantify a priori. First, to make the problem affordable in computational time they capped the magnetic diffusivities to an arbitrary ceiling value and the density to an arbitrary floor value. While these are standard and necessary procedures for common time-integration schemes, they directly alter the heating powers and indirectly affect the electric current density inside the disk. Second, their vertical boundary conditions allow the spontaneous launching of an outflow from the disk. Although outflows are a natural outcome of global magnetized disk models, they suffer from convergence issues in the local shearing box model \citep{fromang13,lesur13} which might affect the energetic balance of the flow. For example, the vertical acceleration of the flow determines the importance of adiabatic cooling. Additionally, to reach a steady state the mass lost through the outflow must be artificially re-injected inside the domain, which supplies internal energy as well if the temperature is left unaltered. 

\section{Summary}

We computed the steady-state vertical structure of a circumstellar disk at radii relevant to planet formation --- $0.5\au$ to $4\au$ in the chosen disk model. We considered weakly ionized and weakly magnetized disks in which angular momentum transport and internal heating are laminar processes, and we precluded outflows. Simplified prescriptions for the gas opacity and ionization fraction allowed us to self-consistently model the energy exchanges in a radiative, non-ideal MHD framework. We considered two different origins for the energy dissipated in the disk, either:
\begin{itemize}
\item the energy of the Keplerian shear is extracted by internal stresses following the growth of the Hall-shear instability inside the disk (labeled `internally-driven' states),
\item or energy is supplied at the surface of the disk through the twisting of a large-scale poloidal field (`externally-driven' states). 
\end{itemize}

Our results support the following conclusions:
\begin{enumerate}
\item Including all three non-ideal MHD effects and a weak poloidal field $\beta > 10^4$, the isothermal and current-free equilibrium can be linearly unstable for midplane ionization fractions as low as $10^{-15}$. The instability saturates in equilibria sustaining angular momentum transport and energy dissipation. Neglecting energetic losses through an outflow, the equivalent `viscosity' coefficients can be as high as $\alpha \sim 10^{-3} - 10^{-2}$.
\item In the absence of linear instability, e.g., at lower ionization fractions, similar levels of dissipation can occur inside the disk if the poloidal field is twisted to equipartition strength with the gas by the disk environment, as expected for global accretion-ejection structures. 
\item The Ohmic and ambipolar resistivities transport energy vertically through the disk, allowing energy thermalization away from its original source. For a sufficiently high opacity $\kappa \gtrsim 10^{-1}\,\mathrm{cm}^2\,\mathrm{g}^{-1}$, the dissipation of electric currents driven from the disk surface can induce order unity temperature variations in the disk interior, and can in some cases lead to convective instability.
\end{enumerate}

We have indications that the 1D equilibria studied in this paper should induce radial inhomogeneities in disks, and that they might be vulnerable to radially dependent perturbations. The steady state spontaneously adopted by three-dimensional flows and the achievable level of dissipation should therefore be pursued with global MHD simulations. Because these issues depend crucially on the gas opacity and ionization fraction, realistic quantitative predictions require a detailed treatment of chemistry and radiation as it becomes computationally affordable \citep{xubai19,wang19,thilesur19}. 

\section*{Acknowledgements}

We thank Gordon Ogilvie and Christian Rab for contextualizing externally-driven solutions and FLD radiative transport in protoplanetary disks. We thank the anonymous reviewer for his careful reading of the paper and for the suggested corrections and clarifications. We also thank Geoffroy Lesur, Xuening Bai and Martin Pessah for their comments on the final draft. William B\'ethune gratefully acknowledges funding by the Deutsche Forschungsgemeinschaft (DFG) through Grant KL 650/31-1, as well as the Isaac Newton Trust and the Department of Applied Mathematics and Theoretical Physics of the University of Cambridge where this project started.



\bibliographystyle{mnras}
\bibliography{biblio} 


\appendix

\section{Alternative example solution} \label{app:example2}

\begin{figure}
\begin{center}
  \includegraphics[width=\columnwidth]{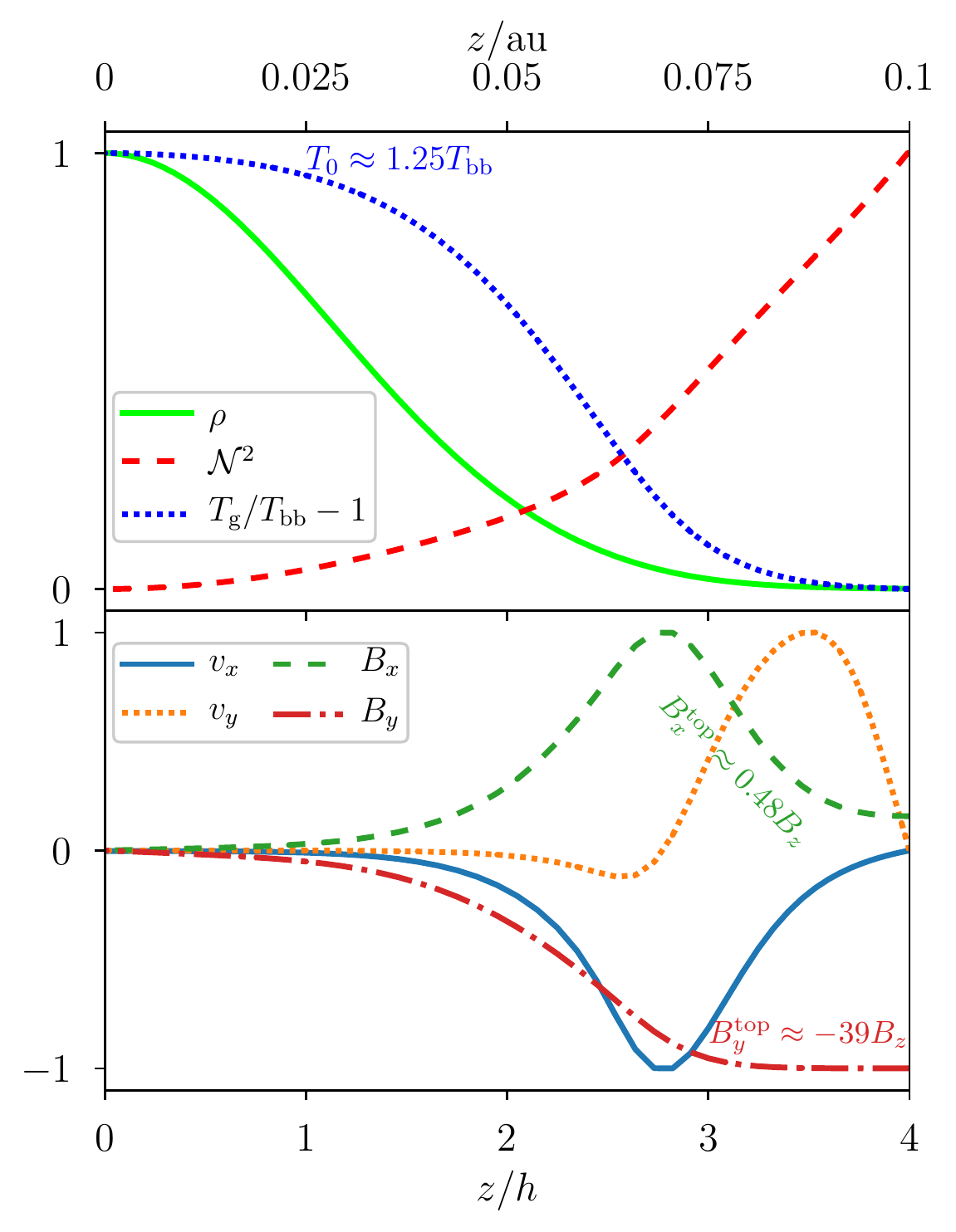}  
  \caption{Vertical profiles of the flow variables in an equilibrium with $\beta=10^6$ at $r=1\au$, normalized by their extremal value for visibility. \emph{Upper panel:} density (solid green), squared \BV frequency (dashed red) and gas temperature relative to $\Tbb$ (dotted blue). \emph{Lower panel:} radial velocity (solid blue), azimuthal velocity (dotted orange), radial magnetic field (dashed green), azimuthal magnetic field (dot-dashed red). \label{fig:mripr1k1b6_fulleq}}
\end{center}
\end{figure}

We draw on Fig. \ref{fig:mripr1k1b6_fulleq} the vertical profiles of an equilibrium with $\beta = 10^6$ at a radius of $r=1\au$. It is twice closer to the star than the reference equilibrium of Sect. \ref{sec:refint}, in a region where the disk is denser and the ionization fraction is lower (see Fig. \ref{fig:ionfraction}). It is threaded by a stronger $B_z$ to reach a comparable temperature contrast. 

The gas temperature is maximal in the midplane with $T_0\approx 1.25 T_{\mathrm{bb}}$, smaller than the reference case shown on Fig. \ref{fig:mripr1k1b7_fulleq}. The squared \BV frequency is also positive everywhere but overall greater than in the reference case, putting this equilibrium further away from convective instability. The velocity and magnetic fields have small amplitudes near the midplane and large amplitudes above $z/h\gtrsim 2$. The azimuthal component of the magnetic field reaches $\Byt\approx -39 B_z$ at the disk surface, about three times weaker then in the reference case. 

\begin{figure}
\begin{center}
\includegraphics[width=\columnwidth]{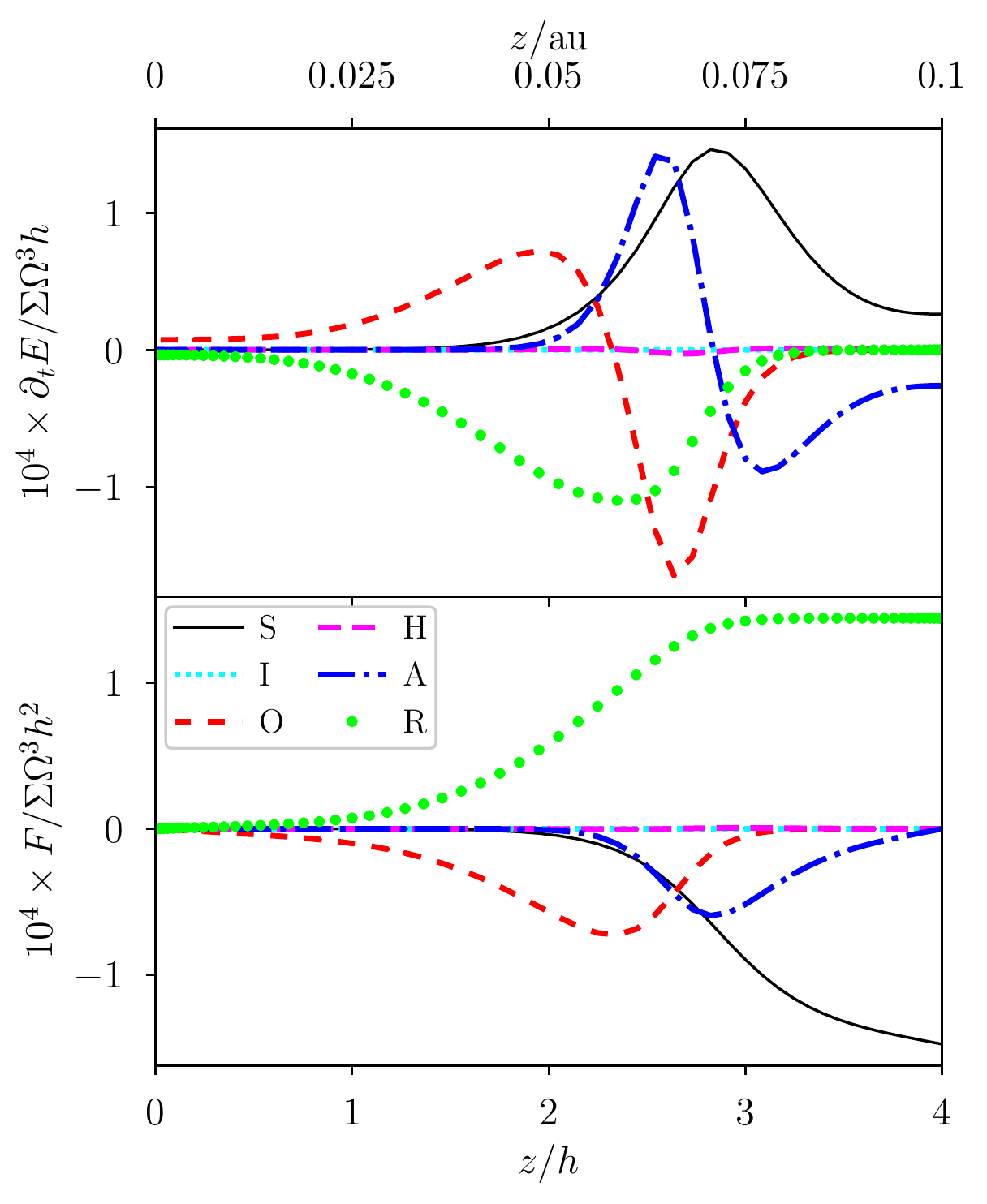}
\caption{Energy budget in the same equilibrium as on Fig. \ref{fig:mripr1k1b6_fulleq}. \emph{Upper panel:} total energy equation \eqref{eqn:nrj}. \emph{Lower panel:} associated energy fluxes, equivalent to the vertical integral of \eqref{eqn:nrj} from the midplane. the different curves correspond to the source term (`S', solid black), the ideal induction (`I', dotted cyan), Ohmic resistivity (`O', dashed red), the Hall drift (`H', dashed magenta), ambipolar diffusion (`A', dot-dashed blue) and radiation (`R', green dots). \label{fig:mripr1k1b6_fullnrj}}
\end{center}
\end{figure}

We decompose the energy equation \eqref{eqn:nrj} for this equilibrium on Fig. \ref{fig:mripr1k1b6_fullnrj}. The source term, related to the extraction of orbital energy via the $xy$ stress, is maximal at $z/h \approx 2.8\pm 0.5$ and essentially zero below $z\lesssim h$. The absence of stress in the midplane indicates that this region is linearly stable to vertical perturbations on a scale $\sim h$. The energy extracted from the shear is transported downward by ambipolar diffusion and then by Ohmic diffusion. The orbital energy can thus be thermalized away from the `active' surface layers which support most of the stress. Even if electric heating predominantly happens above $z/h\gtrsim 1$, this heat accumulates in the midplane for optically thick disks as expressed by \eqref{eqn:conductor} in Sect. \ref{sec:extnrj}.

\section{Test of the implicit integrator} \label{app:integrator}

We verify that our numerical scheme properly captures the physics of the problem by simulating the growth of the Hall-shear instability in a stratified, weakly ionized and isothermal disk. For simplicity we do not include the pressure and radiation equations, and we compute the MHD diffusivities $\eta_{\rm O,H,A}$ assuming a constant ionization fraction $x_e=10^{-12}$. We initialize the disk in a current-free equilibrium \eqref{eqn:current-free} with $\beta=10^6$ and $P=\rho c_s^2$ at all times and altitudes. We add a noise of small amplitude $10^{-6}$ on top of this equilibrium to seed the instability. We keep the damping term $\Omega h^2 \partial_z^2 v_z$ in \eqref{eqn:dtvz} for consistency with the rest of the paper; this term does not affect the linear phase of the instability for which $v_z = 0$.

Separately, we linearize the system of equations \eqref{eqn:DTrho}-\eqref{eqn:DTB} about the current-free equilibrium \eqref{eqn:current-free} assuming an isothermal equation of state $P=\rho c_s^2$. We obtain the set of tangent eigenvalues and eigenmodes numerically. For the equilibrium considered, there are two unstable eigenmodes with associated eigenvalues (growth rates) $2.76\times 10^{-2}\Omega$ and $1.40\times 10^{-1}\Omega$. Starting from the perturbed equilibrium, the fastest growing mode should quickly dominate the vertical structure of the flow. 

\begin{figure}
\begin{center}
\includegraphics[width=\columnwidth]{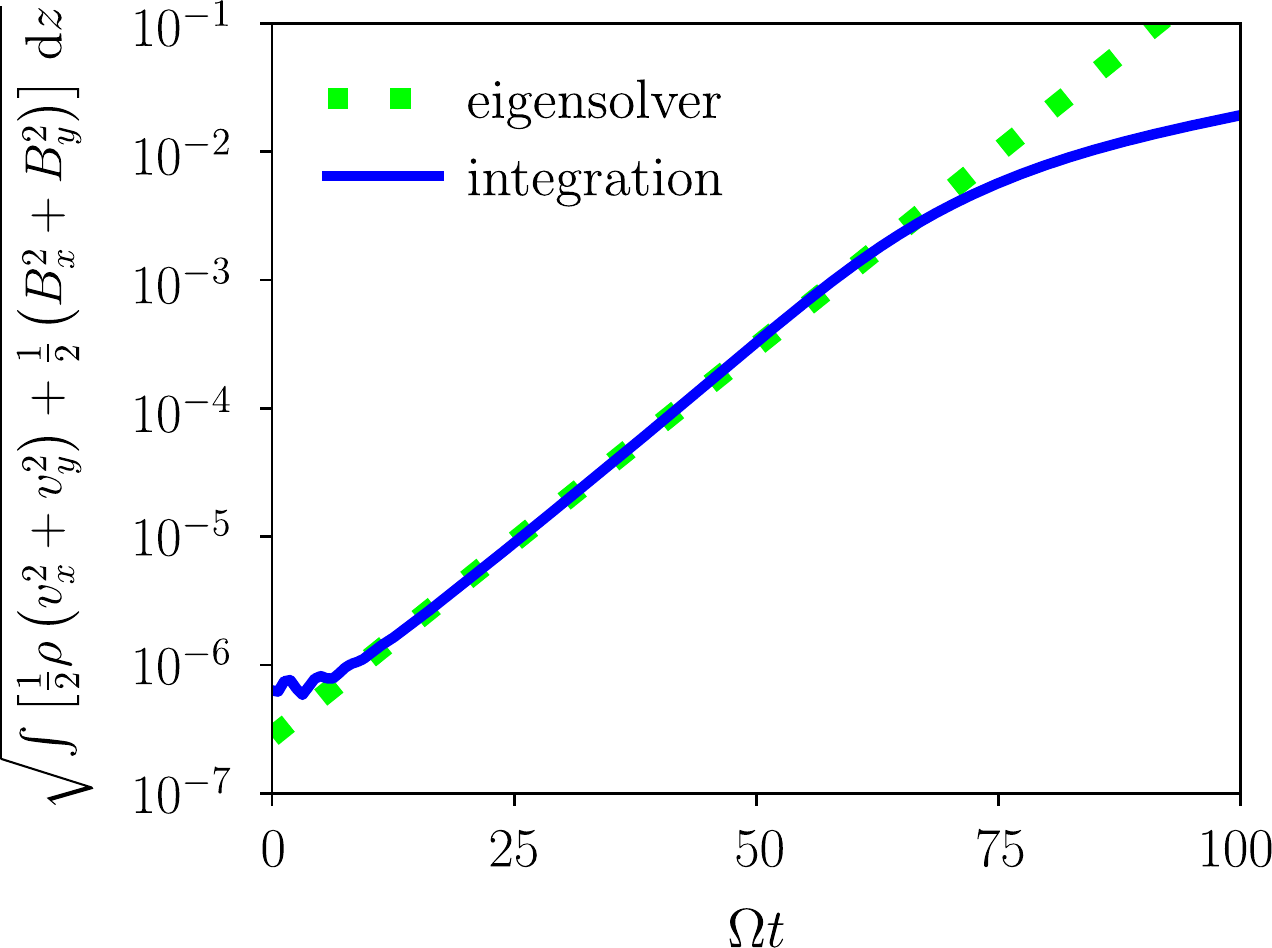}
\caption{Square root of the vertically-integrated energy perturbation when starting from the current-free equilibrium with $\beta=10^6$ and a constant ionization fraction $x_e=10^{-12}$. The solid blue curve corresponds to the first-order implicit integration; the dotted green line indicates the predicted growth rate $1.40\times 10^{-1}\Omega$ of the fastest growing eigenmode. \label{fig:testgrowth}}
\end{center}
\end{figure}

Fig. \ref{fig:testgrowth} shows how the amplitude of the perturbations grows in time. The exponential growth appears after $\Omega t \gtrsim 8$. The growth rate obtained by first-order implicit integration matches the predicted growth rate $1.40\times 10^{-1}\Omega$ to per cent accuracy during this phase. The growth of the perturbations slows down after $\Omega t \gtrsim 60$, marking the non-linear saturation of the instability.

\begin{figure}
\begin{center}
\includegraphics[width=\columnwidth]{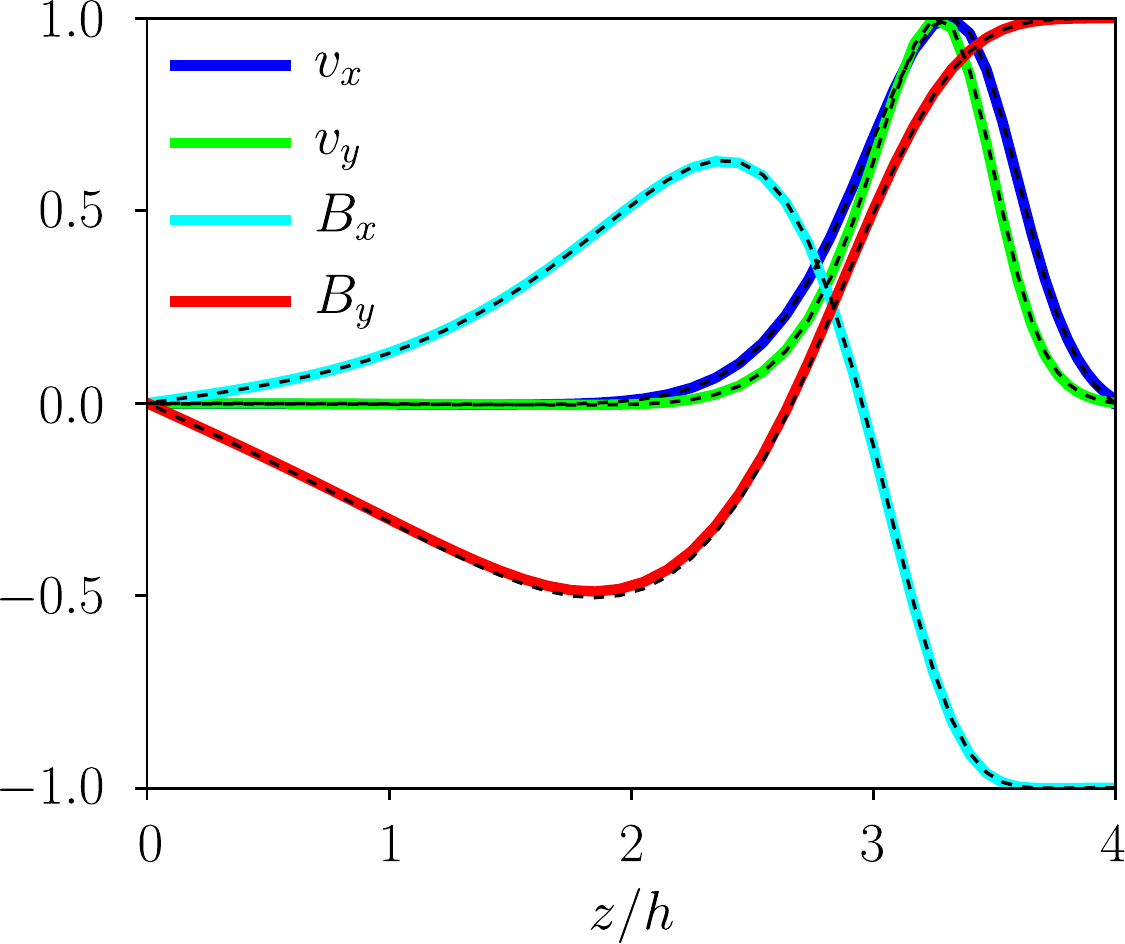}
\caption{Vertical structure of the unstable mode growing in the implicit integration test. The thick solid curves represent the instantaneous profiles at $\Omega t = 35$; the thin dashed curves represent the eigenmode with growth rate $1.40\times 10^{-1}\Omega$ predicted by linear analysis. The curves have been normalized to fit in the range $\left[-1,1\right]$. \label{fig:teststruct}}
\end{center}
\end{figure}

Fig. \ref{fig:teststruct} shows the vertical structure of the flow at $\Omega t = 35$, i.e., during the exponential growth phase. The vertical profiles of velocity and magnetic field match those of the fastest growing eigenmode over the whole extent of the computational domain, confirming that the instability is accurately resolved in space. 

\section{Convergence with domain size} \label{app:convtest}

Throughout this paper we fixed the vertical extent of the domain to $L_z=4h$ so as to focus our resolution near the midplane while including most of the mass and electric current passing through the disk. The gas inside the computational domain is causally connected to the boundaries, so the upper boundary conditions necessarily affect the structure of the steady-state solutions. In particular, imposing $\partial_z B_y=0$ at a finite height might truncate the electric current distribution and lead to under-estimated heat fluxes $Q$. We now examine the convergence rate of the heat flux with domain size $L_z$. 

We vary $L_z$ from $3h$ to $4.5h$ in internally-driven equilibria with $\beta=10^6$ at $r=1\au$. The case $L_z=4h$ corresponds to the equilibrium shown on Fig. \ref{fig:mripr1k1b6_fulleq}. Larger domain sizes $L_z>4.5$ place severe constraints on the numerical stability of the integration scheme when including radiation transport, presumably due to the short radiative time scales of the uppermost layers. Since the energy source is maximal away from the midplane for this choice of $r$ and $\beta$ ($z/h\approx 2.8\pm 1$, cf. upper panel of Fig. \ref{fig:mripr1k1b6_fullnrj}), this equilibrium should be a defavorable case regarding convergence rates. 

\begin{figure}
\begin{center}
\includegraphics[width=\columnwidth]{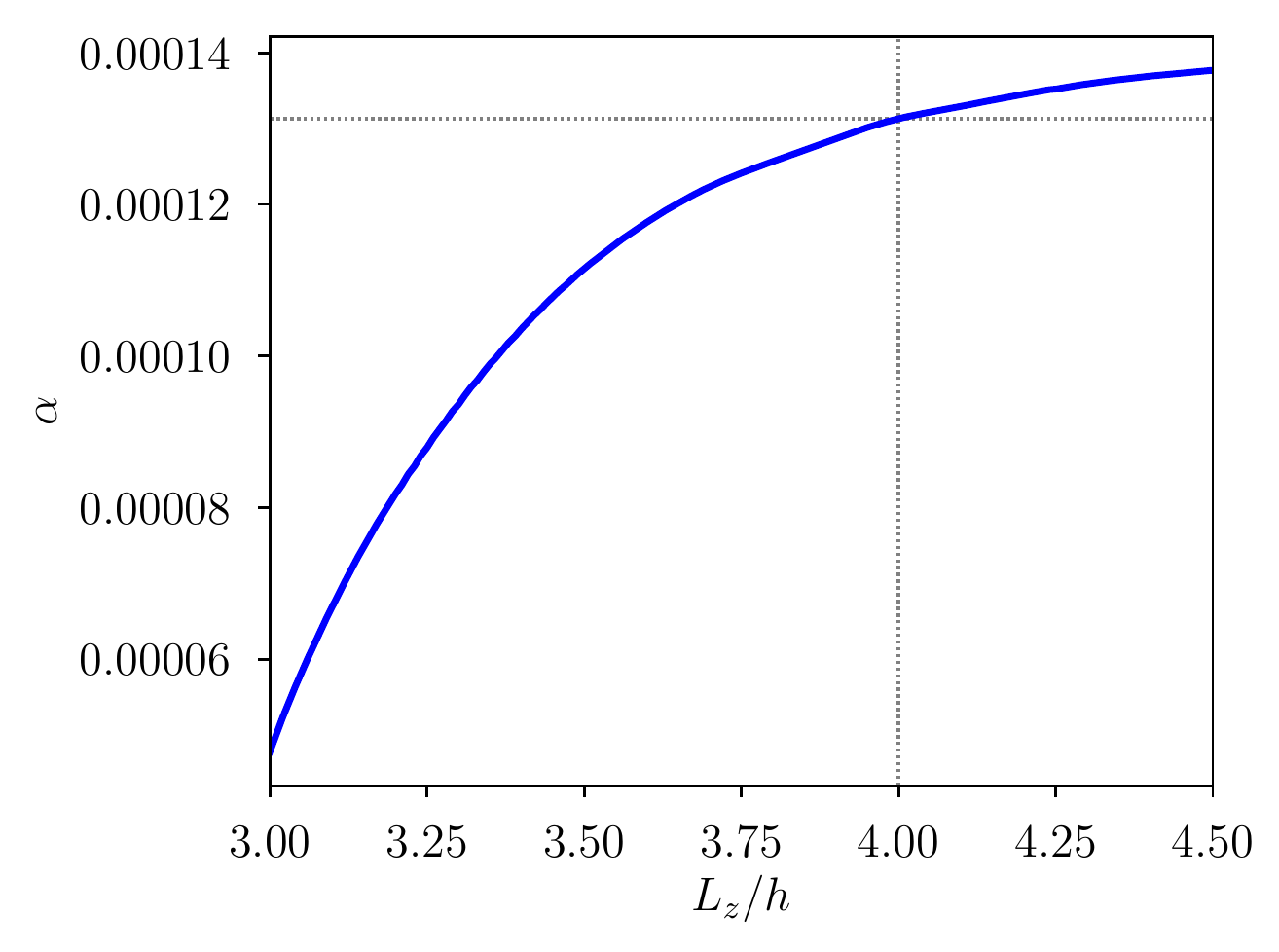}
\caption{Dissipation coefficient $\alpha$ as a function of the box size $L_z$ for internally-driven equilibria with $\beta=10^6$ at $r=1\au$. The dotted lines indicate our standard value for the box size $L_z/h=4$ and the corresponding $\alpha$ as marked on Fig. \ref{fig:mripr1k1_alpha}. \label{fig:convtest}}
\end{center}
\end{figure}

Fig. \ref{fig:convtest} shows that the dissipation coefficient $\alpha$ increases with $L_z$, supporting the idea that the heat flux $Q$ is under-estimated when truncating the domain at a finite height. The heat flux increases mostly below $L_z/h\lesssim 3.5$, corresponding to the range of altitude where energy is extracted from the Keplerian shear by the $xy$ stress, see Fig. \ref{fig:mripr1k1b6_fullnrj}. For $L_z/h\gtrsim 4$ the heat flux keeps slowly increasing with $L_z$, also in agreement with the non-vanishing stress above the disk. Considering the slow convergence rate of $\alpha$ with $L_z$, the dissipation coefficients presented throughout this paper should be placed within ten per cent confidence intervals regarding the sensitivity to the domain size $L_z$.

Our model does not account for the transition from the disk to the stellar environment. A sharp thermo-chemical transition may be induced by stellar X and FUV photons \citep{aresu11,perezchiang11} at altitudes of roughly $z/h \gtrsim 4$, with implications on the ionization fraction and the launching of outflows. Since this transition would affect the plasma conductivity and its energetics, simply extending our 1D disk model to larger domains $L_z \gtrsim 4h$ would not yield more realistic results. 


\bsp	
\label{lastpage}
\end{document}